\pdfoutput=1

\documentclass[11pt,twoside,a4paper,cmspaper,final,collab]{cms-tdr}
\def\svnVersion{491519P}\def\cmsCernNoTag{CERN-EP-2018-313}\def\cmsCernDate{\today}\def\cmsMessage{Published in the European Physical Journal C as \href{http://dx.doi.org/10.1140/epjc/s10052-019-6688-5}{\doi{10.1140/epjc/s10052-019-6688-5.}}}

\begin{document}\cmsNoteHeader{B2G-17-015}

\hyphenation{had-ron-i-za-tion}
\hyphenation{cal-or-i-me-ter}
\hyphenation{de-vices}
\RCS$HeadURL: svn+ssh://svn.cern.ch/reps/tdr2/papers/B2G-17-015/trunk/B2G-17-015.tex $
\RCS$Id: B2G-17-015.tex 487349 2019-01-25 11:09:08Z abenecke $
\newlength\cmsFigWidth
\ifthenelse{\boolean{cms@external}}{\setlength\cmsFigWidth{0.85\columnwidth}}{\setlength\cmsFigWidth{0.4\textwidth}}
\ifthenelse{\boolean{cms@external}}{\providecommand{\cmsLeft}{top\xspace}}{\providecommand{\cmsLeft}{left\xspace}}
\ifthenelse{\boolean{cms@external}}{\providecommand{\cmsRight}{bottom\xspace}}{\providecommand{\cmsRight}{right\xspace}}

\newcommand{\Tp}{\ensuremath{\text{T}}\xspace}
\newcommand{\ZtT}{\ensuremath{\cPZpr \to \cPqt\Tp}}
\newcommand{\Tht}{\ensuremath{\Tp \to \PH\cPqt}}
\newcommand{\Tzt}{\ensuremath{\Tp \to \PZ\cPqt}}
\newcommand{\Twb}{\ensuremath{\Tp \to \PW\cPqb}}
\newcommand{\MZp}{\ensuremath{M_{\cPZpr}}\xspace}
\newcommand{\MTp}{\ensuremath{M_{\Tp}}\xspace}
\newcommand{\rhol}{\ensuremath{\rho_{\mathrm{L}}}\xspace}
\newcommand{\rhoz}{\ensuremath{\rho^0}\xspace}
\newcommand{\gstar}{\ensuremath{\mathrm{G}^*}\xspace}
\newcommand{\ptrel}{\ensuremath{p_{\mathrm{T},\text{rel}}}\xspace}
\newcommand{\chisq}{\ensuremath{\chi^2}\xspace}
\newcommand{\totlumi}{\ensuremath{35.9\fbinv}\xspace}
\newcommand{\Msd}{M_{\text{AK8}}^{\text{SD}}}
\newcommand{\Mrec}{\ensuremath{M_{\cPZpr}^{\text{rec}}}\xspace}
\newcommand{\muf}{\ensuremath{\mu_{\mathrm{F}}}\xspace}
\newcommand{\mur}{\ensuremath{\mu_{\mathrm{R}}}\xspace}
\newcommand{\Hbb}{\ensuremath{\PH_{2\cPqb}}\xspace}
\newcommand{\Hb}{\ensuremath{\PH_{1\cPqb}}\xspace}
\newcommand{\mujets}{\ensuremath{\mu\text{+jets}}}
\newcommand{\ejets}{\ensuremath{\Pe\text{+jets}}}
\newcommand{\ljets}{\ensuremath{\ell\text{+jets}}}
\newcommand{\Wjets}{\ensuremath{\PW\text{+jets}}\xspace}
\newcommand{\Zjets}{\ensuremath{\PZ\text{+jets}}\xspace}
\newcommand{\pp}{\Pp\Pp}
\newcommand{\ZW}{\PZ/\PW}
\newcommand{\jet}{\ensuremath{\mathrm{j}}\xspace}
\newcommand{\pb}{\unit{pb}}
\newlength\cmsTabSkip\setlength{\cmsTabSkip}{1ex}
\providecommand{\cmsTable}[1]{\resizebox{\textwidth}{!}{#1}}
\providecommand{\NA}{\ensuremath{\text{---}}}

\cmsNoteHeader{B2G-17-015}
\title{Search for a heavy resonance decaying to a top quark and a vector-like top quark in the lepton+jets final state in pp collisions at $\sqrt{s} = 13\TeV$}
\titlerunning{Search for a heavy resonance decaying to a top quark and a vector-like top quark at 13\TeV}

\date{\today}

\abstract{ A search is presented for a heavy spin-1 resonance $\cPZpr$
  decaying to a top quark and a vector-like top quark partner $\Tp$ in the
  lepton+jets final state.
  The search is performed using a data set of {\pp} collisions at a
  centre-of-mass energy of 13\TeV corresponding to an integrated
  luminosity of $\totlumi$ as recorded by the CMS experiment at
  the CERN LHC in the year 2016. The analysis is optimised for final states arising from the
  $\Tp$ decay modes to a top quark and a Higgs or \PZ boson ($\Tp
  \to \PH\cPqt, \PZ\cPqt$). The event selection makes use of resolved and
  merged top quark decay products, as well as decays of boosted Higgs bosons and
  \PZ and $\PW$ bosons using jet substructure techniques.  No significant
  deviation from the standard model background expectation is
  observed. Exclusion limits on the product of the
  cross section and branching fraction for $\cPZpr \to \cPqt \Tp, \Tp \to \PH\cPqt, \PZ\cPqt, \PW\cPqb$
  are presented for various combinations of the $\cPZpr$ resonance mass and the vector-like $\Tp$
  quark mass.
  These results represent the most stringent limits to date for the
  decay mode $\ZtT \to \cPqt\PH\cPqt$.
  In a benchmark model with extra dimensions, invoking a heavy spin-1 resonance $\gstar$,
  masses of the $\gstar$ between 1.5 and 2.3\TeV and between 2.0 and 2.4\TeV are excluded
  for $\Tp$ masses of 1.2 and 1.5\TeV, respectively.
}

\hypersetup{%
pdfauthor={CMS Collaboration},%
pdftitle={Search for a heavy resonance decaying to a top quark and a vector-like top quark in the lepton+jets final state in pp collisions at sqrt(s) = 13 TeV},%
pdfsubject={CMS},%
pdfkeywords={CMS, physics, spin-1 resonance, vector-like quarks, extra dimensions}}

\maketitle

\section{Introduction}
\label{sec:introduction}
Many extensions of the standard model (SM) predict the existence of
heavy bosonic resonances, such as the composite spin-1 resonances found in the $\rhoz$ model~\cite{rho}, or
the lightest Kaluza--Klein excitation of the gluon~\cite{gstar} in Randall--Sundrum
models~\cite{extradimension1,Randall:1999vf}.
In models that invoke such states to address the hierarchy problem, these resonances are
required to cancel top quark loop contributions to the radiative corrections that
would otherwise drive the Higgs boson mass up to the Planck scale.
As a consequence, these resonances typically feature an enhanced coupling
to third-generation SM quarks, resulting in decays predominantly to SM top quarks.
The ATLAS and CMS Collaborations have performed
searches for such resonances in their proton-proton ($\pp$) collision data sets at
centre-of-mass energies $\sqrt{s} = 8$~\cite{ttbar_atlas, Chatrchyan:2013lca, Khachatryan:2015sma}
and 13\TeV~\cite{Aaboud:2018mjh, ttbar, Sirunyan:2018ryr},
leading to stringent exclusion limits on the product of the resonance cross section and branching fraction,
 and therefore also on the resonance masses when interpreted in the context of specific models.
 However, in some models describing physics beyond the standard model (BSM) the new heavy resonance state
is accompanied by an additional fermionic sector realised as a
non-chiral (or vector-like) fourth generation of quarks.
Vector-like quarks (VLQs) are fermions whose left- and right-handed
components transform in the same way under the electroweak symmetry
group. For this reason, direct mass terms for VLQs are not forbidden in the
BSM Lagrangian. Furthermore, unlike sequential fourth-generation chiral quarks,
the existence of VLQs is not yet excluded by recent Higgs boson measurements~\cite{handbook}.

\begin{figure}[htb]
\centering
\includegraphics[width=.4\textwidth]{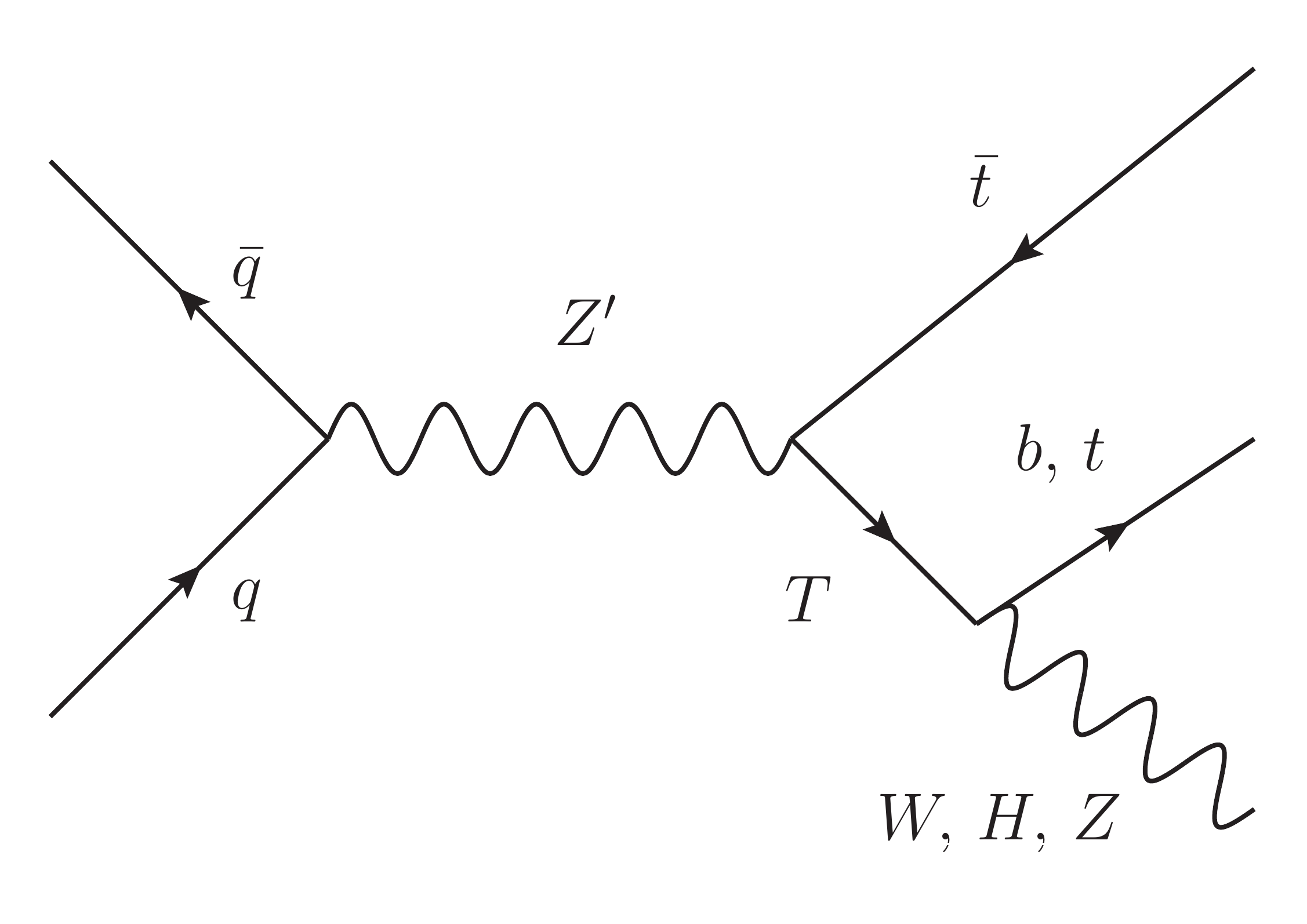}
\caption{Leading order Feynman diagram for the production of a spin-1 resonance $\cPZpr$ and its decay, along with the possible decays of the vector-like quark \Tp.}
\label{fig:ZPrime}
\end{figure}

Examples of BSM models with these additional particles are models
with a heavy gluon~\cite{gstar}, a composite Higgs
boson~\cite{composit2,Barducci:2012kk, composit1, Barducci:2015vyf},
or extra spatial dimensions~\cite{extradimension1,extradimension2}. In these models, decays of
the heavy resonance into final states with VLQs are allowed. The analysis
presented in this paper searches for evidence of the production of a
heavy spin-1 resonance, denoted as $\cPZpr$, with decays to an SM top quark and a vector-like top
quark partner $\Tp$, $\ZtT$. This decay mode is dominant
for the intermediate $\cPZpr$ mass region
in which the decay is kinematically allowed ($M_{\cPqt}+\MTp<\MZp$) and the
decay to two VLQ top quark partners ($\cPZpr \to \Tp\Tp$) is kinematically
forbidden ($\MZp<2\MTp$).
The VLQ $\Tp$ considered in this analysis has three decay channels:
to a Higgs boson and a top quark $(\Tht)$,
to a \PZ boson and a top quark $(\Tzt)$, and to a $\PW$ boson and a bottom quark $(\Twb)$.
The leading-order Feynman diagram showing the production
mode of the $\cPZpr$ boson and its subsequent decay, including the decay of the VLQ, is shown in
Fig.~\ref{fig:ZPrime}.

A first search for the production of a heavy $\cPZpr$ resonance decaying
to $\cPqt\Tp$ was performed by the CMS Collaboration~\cite{emanuele} using
a data set corresponding to an integrated luminosity of 2.6\fbinv, recorded at a centre-of-mass energy of
$\sqrt{s}=13\TeV$. The analysis was optimised for the decay mode $\Twb$ and made use of
the all-hadronic final state, where both the top quark and $\PW$ boson are highly Lorentz-boosted,
resulting in a three-jet event topology. No deviation from the SM
expectation was observed and upper limits on the cross section were obtained, ranging
from 0.13 to 10\pb depending on the masses of the $\cPZpr$ and $\Tp$.
A search for the single production of a vector-like quark $\Tp$ decaying to a \PZ boson and a
top quark in the dilepton+jets final state has also been performed,
with an interpretation of the results in the context of a $\cPZpr$ decaying into $\cPqt\Tp$~\cite{singleVLQ}.
Upper limits on the cross section ranging from 0.06 to 0.13\pb were obtained,
for the production of a $\cPZpr$ with the subsequent decays to $\cPqt\Tp$ and $\Tzt$.
Searches for $\Tp$ pair production have been performed by the ATLAS~\cite{Aaboud:2017qpr,
Aaboud:2017zfn, Aaboud:2018xuw, Aaboud:2018saj, Aaboud:2018wxv, Aaboud:2018pii}
and CMS~\cite{Sirunyan:2017pks, Sirunyan:2018omb} Collaborations, placing a lower bound
of $\approx$1.3\TeV on the VLQ mass.

The analysis presented here is optimised for $\ZtT$ with the $\Tp$
decay modes $\Tzt$ and $\Tht$,
in the monolepton+jets final state.
The search is performed in a data set of {\pp} collisions at
$\sqrt{s}=13\TeV$ corresponding to
$\totlumi$~\cite{CMS-PAS-LUM-17-001} as recorded by the CMS experiment
during the year 2016.

The two decay channels ($\ZtT \to \cPqt\PH\cPqt$,
$\ZtT \to \cPqt\PZ\cPqt$) each produce two top quarks accompanied by one boson.  The event selection in the single lepton+jets final state
relies on the leptonic decay of one of these top quarks. The {\PH} or {\PZ} boson from the $\Tp$ decay
 is expected to receive a large Lorentz boost
because of the large mass difference between the $\Tp$ and the
boson. The resultant hadronic decay products will be merged
and are thus reconstructed as a single broad jet.
 Jet substructure techniques are used for the
boson identification~\cite{CMS-PAS-JME-16-003} and in the
event categorisation. Additional categories for both resolved and merged
decays of the other hadronically decaying top quark are considered to
ensure a high sensitivity over a broad range of the $\cPZpr$ mass.  In
all categories, the $\cPZpr$ mass is reconstructed
by considering various combinations of reconstructed objects, with the final combination determined by a
$\chisq$ metric.

Limits at 95\% confidence level (\CL) are derived for all three $\Tp$ decay
channels ($\Tht$, $\Tzt$, $\Twb$) using a template-based statistical
evaluation of the reconstructed mass spectra of the $\cPZpr$ boson from all
categories.  A mass range of the $\cPZpr$ from 1.5 to 4.0\TeV
and of the $\Tp$ from 0.7 to 3.0\TeV is considered.  The
rate of the dominant SM background processes ($\ttbar$ and
$\Wjets$) predicted by simulation is constrained using the mass spectra in dedicated control regions that
enhance these background processes, and are fit simultaneously with the signal regions.

The search is performed in a model-independent manner by scanning over a large
range of possible masses of the $\cPZpr$ and $\Tp$ and couplings of the $\Tp$ to various final states.
The results are then interpreted in the context of two theory benchmark models.

In the $\gstar$ model~\cite{gstar}, ten new VLQs
(\Tp, \PB, $\tilde{\Tp}$, $\tilde{\PB}$, $\Tp_{5/3}$, $\Tp_{2/3}$,
$\Tp^\prime$, $\PB^\prime$, $\PB_{-1/3}$, $\PB_{-4/3}$) are predicted with
well-defined relationships between their masses. In this analysis, the $\Tp$
mass is varied, whilst other masses are related by
$M_{\Tp_{5/3}} = M_{\Tp_{2/3}} = M_{\Tp} \cos \phi_{\mathrm{L}}$.
The mixing angle $\cos \phi_{\mathrm{L}}$ governs the degree of compositeness of
the left-handed quark doublet $(\cPqt_{\mathrm{L}}, \cPqb_{\mathrm{L}})$, and hence the
relative coupling of the lightest spin-1 Kaluza--Klein excitation of the gluon,
$\gstar$, to third-generation quarks compared to the other two generations of quarks.
A benchmark scenario with parameters $\tan \theta = 0.44$, $\sin \phi_{\mathrm{tR}} = 0.6$, and $Y_{*} = 3$
is used in this analysis, leading to $\cos \phi_{\mathrm{L}} = 0.84$.
A description of the benchmark and its parameters can be found in Ref.~\cite{gstar}.
In this model the branching fractions ($\mathcal{B}$) of the
$\Tp$ decay to $\PW \cPqb$, $\PH \cPqt$, and $\PZ \cPqt$ are chosen to be
0.5, 0.25, and 0.25, respectively.

The $\rhoz$ model~\cite{rho} predicts a heavy spin-1 resonance, $\rho$, along with a
multiplet of four new vector-like
quarks, with two of the vector-like quarks ($\Tp$, $\text{B}$) representing the heavy
partners of the top and $\cPqb$ quarks, respectively. Other
exotic vector-like quarks are also predicted: $\text{X}_{2/3}$ with a charge of
$2\,e/3$, and $\text{X}_{5/3}$ with a charge of $5\,e/3$,
where $e$ is the magnitude of the charge of the electron.
A benchmark scenario with parameters $y_{\mathrm{L}}=c_2=c_3=1$ and $g_{\rhol}=3$
is used in this analysis, where a description of the benchmark and its parameters can be
found in Ref.~\cite{rho}.
In this model the branching fractions of the
$\Tp$ decay to $\PW \cPqb$, $\PH \cPqt$, and $\PZ \cPqt$ are chosen to be
0, 0.5, and 0.5, respectively.

After a short description of the CMS experiment
(Section~\ref{sec:detector}), the reconstruction of events is discussed in
Section~\ref{sec:reconstruction}, and the data sets and the simulated
samples are introduced in Section~\ref{sec:datasets}. The event selection,
the categorisation, and the procedures for the mass reconstruction are
discussed in Section~\ref{sec:sel}.  In Section~\ref{sec:background} the
background estimation procedure is explained, and
Section~\ref{sec:systematics} gives an overview of the systematic
uncertainties. Finally, the search results and the interpretation in
the benchmark models are presented in
Section~\ref{sec:results}.
This paper concludes with a summary in Section~\ref{sec:summary}.

\section{The CMS detector}
\label{sec:detector}
The central feature of the CMS apparatus is a superconducting solenoid
of 6\unit{m} internal diameter, providing a magnetic field of
3.8\unit{T}. Contained within the solenoid volume are
a silicon pixel and strip tracker, a lead tungstate crystal
electromagnetic calorimeter (ECAL), and a brass and scintillator
hadron calorimeter (HCAL), each composed of a barrel and two endcap
sections. Muons are detected in gas-ionization chambers embedded in
the steel flux-return yoke outside the solenoid. Extensive forward
calorimetry complements the pseudorapidity ($\eta$) coverage provided by the barrel and endcap
detectors. Events of interest are selected using a two-tiered trigger
system~\cite{Khachatryan:2016bia}.
The first level, composed of custom hardware processors,
uses information from the calorimeters and muon detectors to select events at
a rate of around 100\unit{kHz} within a time interval of less than 4\mus.
The second level, known as the high-level trigger, consists of a
farm of processors running a version of the full event reconstruction
software optimised for fast processing, and reduces the event rate to
around 1\unit{kHz} before data storage.
A more detailed description of the CMS detector, together
with a definition of the coordinate system used and the relevant
kinematic variables, can be found in Ref.~\cite{Chatrchyan:2008aa}.

\section{Event reconstruction}
\label{sec:reconstruction}
The particle-flow (PF) algorithm~\cite{Sirunyan:2017ulk} deployed by the CMS Collaboration
aims to reconstruct and identify each individual particle in an event,
 with an optimised combination of information from the various elements of the CMS
detector. The energy of photons is obtained from the ECAL
measurement. The energy of electrons is determined from a combination of the electron
momentum at the primary interaction vertex as determined by the tracker, the energy of
the corresponding ECAL cluster, and the energy sum of all bremsstrahlung photons
spatially compatible with originating from the electron track.
The energy of muons is obtained from the curvature of the corresponding track.
The energy of charged hadrons is determined from a combination of their momentum measured
in the tracker and the matching ECAL and HCAL energy deposits, corrected for zero-suppression
effects and for the response function of the calorimeters to hadronic showers. Finally,
the energy of neutral hadrons is obtained from the corresponding corrected ECAL and HCAL energies.

The primary $\pp$ interaction vertex is taken to be the reconstructed vertex
with the largest value of summed
physics-object $\pt^2$, where \pt is the transverse momentum.
Here the physics objects are the objects returned by a jet finding
algorithm~\cite{Cacciari:2008gp,Cacciari:2011ma} applied to all
tracks associated with the vertex, plus the negative vector sum of the \pt of those jets.

Muons are reconstructed through a fit to hits in both the inner
tracking system and the muon spectrometer~\cite{Chatrchyan:2012xi,Sirunyan:2018}.
Muons must satisfy identification and reconstruction requirements on the impact
parameters of the track, the number of hits reconstructed in both the
silicon tracker and the muon detectors, and the uncertainty in the
\pt.  These quality criteria ensure a precise measurement of the
four-momentum, and rejection of badly reconstructed muons.

Electron candidates are required to have a match between the energy
deposited in the ECAL and the momentum determined from the reconstructed
track~\cite{Khachatryan:2015hwa}.  To suppress the multijet background,
electron candidates must pass stringent identification
criteria.  These include
requirements on the geometrical matching between ECAL deposits and
position of reconstructed tracks, the ratio of the energies deposited
in the HCAL and ECAL, the spatial distribution of the ECAL depositions, the
impact parameters of the track, and the number of reconstructed hits
in the silicon tracker.

In the $\cPZpr$ signal targeted by this analysis, the lepton is emitted in the
decay chain of a top quark ($\cPqt \to \cPqb \ell \nu_{\ell}$) at high \pt.
Because of the Lorentz boost of the top quark, the lepton is expected to be in
angular proximity to a $\cPqb$ quark, and therefore conventional lepton isolation
criteria would lead to a loss in signal efficiency.  Instead, a dedicated
two-dimensional criterion is used to reduce the background of leptons
arising from heavy-flavour quark decays in multijet events produced
through the strong interaction.
This criterion is discussed in Section~\ref{sec:preselection}.

In this analysis hadronic jets are reconstructed from PF candidates
using the anti-\kt algorithm~\cite{Cacciari:2008gp} as implemented in the
\FASTJET software package~\cite{Cacciari:2011ma}. Since the
analysis targets both resolved and merged top quark decays, jets are
clustered with two values of the distance parameter $R$; $R = 0.4$
(AK4 jets) for the reconstruction of resolved top quark decay
products, and $R = 0.8$ (AK8 jets) for the reconstruction of
merged $\PW$, \PZ, and Higgs boson decay products as well as merged top quark
decays.  In the jet clustering procedure, charged PF candidates associated with
non-primary vertices are excluded. The jet
momentum is determined as the vectorial sum of all particle momenta in
the jet, and is found from simulation to be within 5 to 10\% of the
true momentum over the whole \pt spectrum and detector acceptance.  A
correction based on the area of the jet, projected on the front face
of the calorimeter, is used to correct for the extra energy
clustered in jets due to additional inelastic proton-proton interactions
within the same or adjacent bunch crossings (pileup)~\cite{Cacciari:2011ma}.
Jet energy corrections are
derived from simulation in order to bring the measured response of jets to
that of particle level jets on average. Dijet, multijet, photon+jet,
and leptonically-decaying {\PZ}+jet events are used to perform in situ
measurements of the momentum balance to derive corrections for
residual differences in jet energy scale in data and simulation~\cite{Khachatryan:2016kdb}.
Additional selection criteria are
applied to each event to remove spurious jet-like features originating
from isolated noise patterns in certain HCAL regions~\cite{CMS-DP-2016-061}.  The clustered
jets also contain leptons.  To avoid double counting of the lepton momentum in an
event, the lepton used for the reconstruction of the $\PW$ boson from the
leptonic top quark decay is removed from an AK4 jet if the lepton overlaps with the jet
within the jet's radius parameter, $\Delta R(\ell,\, \jet) < 0.4$,
where $\Delta R (\ell,\, \jet) = \sqrt{\smash[b]{[\Delta\eta(\ell,\, \jet)]^2+[\Delta\phi(\ell,\, \jet)]^2}}$,
and $\Delta\eta(\ell,\, \jet)$ and $\Delta\phi(\ell,\, \jet)$ are the separations in pseudorapidity and
azimuthal angle, respectively, between the lepton and jet.
The momentum of the lepton is subtracted from that of the jet before
jet energy corrections are applied.
Larger radius AK8 jets that overlap with the lepton within $\Delta R(\ell,\, \jet) < 0.8$ are not considered in this analysis.
Jets that are produced by the decay of $\cPqb$ quarks
can be identified using the combined
secondary vertex discriminator~\cite{Sirunyan:2017ezt}.
An AK4 jet is denoted as being $\cPqb$ tagged if it fulfils the medium working point of
the discriminator, which has an efficiency of 63\% for correctly
identifying a $\cPqb$ quark jet, with a 1\% probability of misidentifying a
light-flavour jet as $\cPqb$ tagged (a mistag).

The boosted bosons and merged top quark decays are identified by applying so-called taggers to AK8 jets.
Each tagger requires the jet
mass to be within a certain range, along with additional criteria on
substructure variables such as N-subjettiness~\cite{Thaler:2010tr} or
subjet $\cPqb$ tagging~\cite{Sirunyan:2017ezt}.  The
jet mass is computed after applying a modified mass-drop
algorithm~\cite{jetmass_nnll1,Butterworth:2008iy}, known as the
\emph{soft drop} algorithm~\cite{Larkoski:2014wba}, which eliminates
soft, large-angle radiation resulting from SM quantum chromodynamics (QCD) processes.
This improves the jet mass resolution for the
reconstructed boson and top quark. It also reduces the mass of jets initiated
by single quarks or gluons, thereby improving discrimination between jets from
true boson or top quark decays, and those from background QCD multijet events.
Furthermore, it helps mitigate the effect of pileup~\cite{CMS-PAS-JME-16-003}.
The N-subjettiness variable $\tau_{\mathrm{N}}$ quantifies the compatibility
of the jet clustering with the hypothesis that exactly N subjets are present,
with small values of $\tau_{\mathrm{N}}$ indicating greater compatibility.
The N-subjettiness ratios $\tau_{21} = \tau_2 / \tau_1$ and $\tau_{32} =
\tau_3 / \tau_2$ are calculated prior to the application of the soft drop algorithm, and are used
to reject background jets arising from the
hadronization of single quarks or gluons.  Jets from hadronic $\ZW$ boson decays
in signal events are characterized by smaller values of $\tau_{21}$
in comparison to jets from QCD multijet background processes, and
similarly jets from merged hadronic top quark decays have smaller values of
$\tau_{32}$ than background jets.  For each AK8 jet, two subjets are obtained from the
soft drop algorithm.  An AK8 jet can have up to two subjet $\cPqb$ tags
depending on how many subjets fulfil the loose working point of the
$\cPqb$ tagging discriminator. In contrast to the medium working point applied to AK4 jets, the loose
working point has a larger $\cPqb$ tagging efficiency of 83\%, at the expense of a
larger mistag probability of 9\%.

The missing transverse momentum \ptmiss is defined as the magnitude of
the vector sum of the transverse momenta of the reconstructed PF
objects, $\ptvecmiss$.  The value of \ptmiss is modified to account for corrections to
the energy scale of the reconstructed AK4 jets in the event.

\section{Data and simulated samples}
\label{sec:datasets}
The analysis is based on the data set of $\pp$ collisions recorded by
the CMS detector during the year 2016. Events targeting the decay of a
top quark to a final state including a muon are selected with a
high-level single-muon trigger that requires the
presence of at least one muon candidate with $\pt>50\GeV$ and $\abs{\eta}<2.4$. For events targeting a final state with an electron,
the high-level trigger requires the presence of at least one electron candidate with $\pt>115\GeV$
and $\abs{\eta}<2.5$, or at least one photon with $\pt>175\GeV$ and $\abs{\eta}<2.5$.
The latter requirement ensures events containing electrons with a high \pt are efficiently selected,
as the requirements on ECAL shower shapes are less stringent for photons than for electrons.
Given the highly boosted topology of the final-state
objects, no isolation requirements are applied to the lepton
candidates at the trigger level. The electron trigger threshold is significantly
larger than the muon trigger threshold,
 since the non-isolated electron trigger selects a large number of hadrons incorrectly
 identified as electrons.
Both recorded data sets correspond to
an integrated luminosity of $\totlumi$~\cite{CMS-PAS-LUM-17-001} .

The spin-1 resonance signal samples are generated with the leading-order (LO)
mode of \MGvATNLO 2.2.2~\cite{Alwall:2014hca} as a high mass
resonance with SM-like couplings using the $\gstar$ model~\cite{gstar}. The \PYTHIA
8.212~\cite{pythia82} event generator with the CUETP8M1 underlying event
tune~\cite{Skands:2014pea,Khachatryan:2015pea}
is used to model the parton showering and underlying event. Separate samples
for the different decay channels of the $\Tp$ are produced, so that each sample
has a branching fraction of 100\% to the chosen decay channel.
Throughout this paper, a generic spin-1 heavy resonance will be referred to as $\cPZpr$,
whilst interpretations within a given model will refer to their specific resonance names.

We consider only $\cPZpr$ and $\Tp$ masses that result in a significant branching
fraction of the $\cPZpr$ to $\cPqt\Tp$. Scenarios where the mass of the $\cPZpr$ is smaller
than the sum of the top quark mass and the $\Tp$ mass are not considered,
since the $\cPZpr$ would then decay to SM quark pairs, and such scenarios have
been largely excluded by previous searches~\cite{ttbar,ttbar_atlas}.
Masses of the $\cPZpr$ larger than twice the $\Tp$ mass are also
not considered, as in such cases the $\cPZpr$ decays predominantly to $\Tp$ pairs,
resulting in a large $\cPZpr$ width. Such large masses are better targeted by direct
searches for $\Tp$ pair production.

Two values of the $\cPZpr$ width are considered, corresponding to 1\% or 30\% of its mass.
The $\Tp$ width is set to 1\% of its mass.
For the $\cPZpr$ and $\Tp$ mass parameter space considered in this analysis
the total $\cPZpr$ decay width in the two considered theoretical models is
always less than 20\% of its mass.
Since the experimental resolution is approximately 15\%, the samples with the $\cPZpr$ width
set to 1\% are dominated by the experimental resolution, and
are thus used in the interpretation of the results.
The samples generated with the width of 30\% are used as cross-checks and help to
confirm that the conclusions do not change for scenarios with $\cPZpr$
widths somewhat larger than the experimental resolution for high masses of the $\cPZpr$.
Furthermore, it was checked that scenarios with $\Tp$ widths of up to 30\%, with a $\cPZpr$ width equal to or larger than that of the $\Tp$, do not
significantly affect the resolution of the $\cPZpr$ mass, and therefore the experimental limits obtained
with the $\Tp$ width set to 1\% are also valid for larger $\Tp$ width scenarios.

The $\gstar$ model considers only left-handed $\Tp$ quarks. The $\rhoz$ model also
allows for a right-handed $\rho_{\mathrm{R}}$ coupling to $\Tp$ quarks. For the $\Tht$ decay mode the kinematic
distributions in the $\gstar$ model and $\rhoz$ model are the same. While for the
$\Tzt$ and $\Twb$ decay modes the $\ZW$ boson \pt spectra are similar for the left-handed
$\rhol$ and the $\gstar$, the ratio of the distributions for left- and
right-handed scenarios in the $\rhoz$ model deviates from unity by up to 30\%.
In this analysis only the decays of the left-handed $\rhol$ are considered.

Simulated event samples for the SM background processes Drell--Yan (DY)+jets,
also referred to as $\Zjets$, and $\Wjets$ are computed at next-to-leading-order
(NLO) precision in QCD with \MGvATNLO.  The parton
showering is calculated using \PYTHIA 8 following the FxFx merging
scheme~\cite{fxfx}.  Background events from QCD multijet processes are simulated
using \PYTHIA 8.  For the simulation of the underlying event, the tune
CUETP8M1 is used in \PYTHIA 8 for the $\Wjets$, $\Zjets$, and QCD multijets samples.

The simulation of SM $\ttbar$ and single top quark (ST) background events is
performed with the {\POWHEG} event generator~\cite{Frixione:2007nw,Campbell:2014kua,Alioli:2009je,Re:2010bp,Nason:2004rx,Frixione:2007vw,Alioli:2010xd,Alioli:2011as,Frederix:2012dh},
using {\POWHEG} v1.0 for the simulation of $\cPqt\PW$ events,
whilst {\POWHEG} v2.0 was used for the simulation of {\ttbar} and all other single top quark processes.
The \PYTHIA 8 generator was used for the showering in both versions of \POWHEG.
An observed discrepancy between simulation and data in the
top quark \pt spectrum is corrected with a reweighting procedure based on
measurements of the top quark \pt
spectrum~\cite{Sirunyan:2017mzl,Khachatryan:2016mnb}.  The
underlying event tune CUETP8M2T4~\cite{CMS-PAS-TOP-16-021} is used in \PYTHIA 8 for the $\ttbar$ and
single top quark samples.

All events are generated with the NNPDF 3.0 parton distribution
functions (PDFs)~\cite{Ball:2014uwa}. The detector response is
simulated with the {\GEANTfour} package ~\cite{Agostinelli:2002hh}.
Simulated events are processed through the
same software chain as used for collision data.
All simulated event samples include the simulation of pileup, and are reweighted to
match the observed distribution of the number of pileup interactions
in data.

\section{Event selection, categorisation, and mass reconstruction}
\label{sec:sel}
\subsection{Event selection}
\label{sec:preselection}

All events are required to contain at least one reconstructed interaction vertex
within a volume 24\cm in length and 2\cm in radius, centred on the mean $\pp$ collision
point~\cite{Chatrchyan:2014fea}.

Since there are differences in the way the electrons and muons from top quark decays are treated,
the analysis considers the {\ejets} and {\mujets} channels separately.
In the {\mujets} channel exactly one reconstructed muon with $\pt >
53\GeV$ and $\abs{\eta} < 2.4$ is required. In the {\ejets} channel
exactly one electron with $\pt > 125\GeV$ and $\abs{\eta} < 2.5$ is
required.  Events with an electron candidate located inside the
transition region between the ECAL barrel and endcaps ($1.44 <
\abs{\eta} <1.57$) are rejected. In the {\ejets} channel, an
additional requirement of $\ptmiss > 90 \GeV$ from the associated neutrino
is introduced to reduce the background of hadrons falsely identified as electrons
in QCD multijet events.

Because of the boosted nature of the signal, conventional lepton
isolation criteria are not applicable. Instead, in
both the {\ejets} and {\mujets} channels, events are required to pass
a two-dimensional selection of either $\Delta R (\ell,\, \jet) > 0.4$ or
$\ptrel(\ell,\, \jet) > 40\GeV$, where $\jet$ is the AK4 jet with
minimal angular separation $\Delta R$ from the lepton $\ell$ (electron or
muon), and $\ptrel(\ell,\, \jet)$ is the component of the lepton
momentum orthogonal to the axis of jet $\jet$. Only AK4 jets with $\pt
>15\GeV$ are considered in this criterion. 
The chosen working point has an 
efficiency of ${\approx}30\%$ for a lepton with $\pt = 100\GeV$, increasing with \pt
and reaching a plateau of ${\approx}90\%$ at $\pt=400\GeV$.
The background rejection rate is 99.5\% at $\pt = 100\GeV$,
decreasing to ${\approx}94\%$ at $\pt = 400\GeV$.

In order to reconstruct the boosted $\PH$/\PZ/$\PW$ boson or merged top quark decays,
events are required to contain at least one AK8 jet with
$\pt > 250\GeV$ and a soft drop jet mass $\Msd
> 30\GeV$.

\subsection{Event categorisation}

\begin{figure}[tb]
\centering
\includegraphics[width=0.49\textwidth]{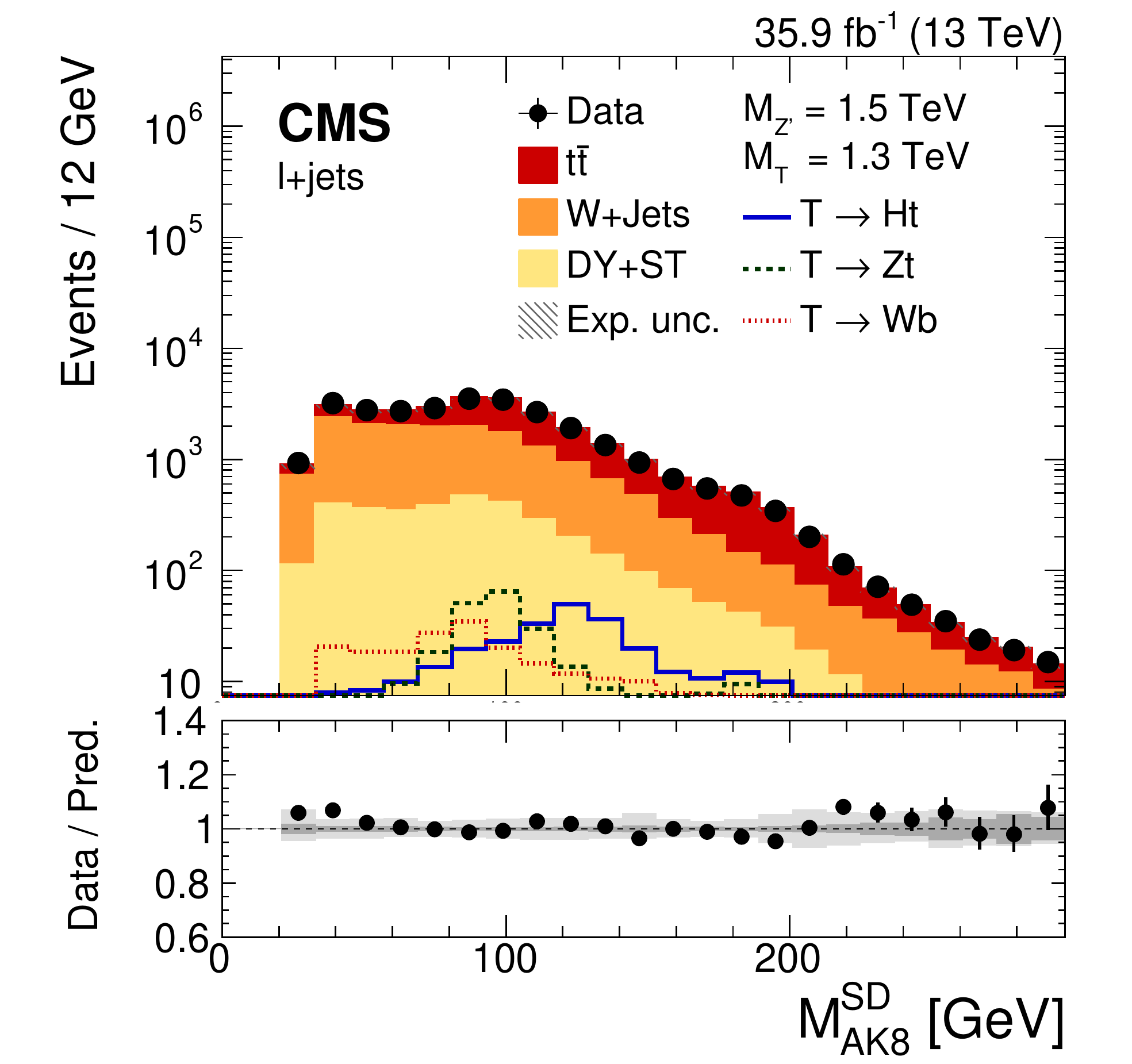}
\caption{Distribution of the soft drop mass of jets as reconstructed
  with the anti-\kt jet algorithm with $R=0.8$ after the event selection.
  Events are shown in the combined lepton+jets channel, with contributions
  from data, simulated signal samples, and the simulated SM backgrounds without corrections from the fit to data (explained in Section~\ref{sec:background}).
  The expected signal distribution from various $\Tp$ decay modes is
  shown for the example mass configuration $\MZp=1.5\TeV$ and
  $\MTp=1.3\TeV$ with a nominal cross section $\sigma(\cPZpr \to \cPqt\Tp)$ of 1\pb.
  The lower panel shows the ratio of data to predicted background. Here the darker grey band indicates the statistical uncertainty, whilst the lighter grey band shows the combined statistical and systematic uncertainty.}
\label{fig:jetmass}
\end{figure}

After the event selection, different taggers are used for the
identification of hadronic decays of boosted $\ZW$ bosons, Higgs bosons,
and top quarks, called in the following $\ZW$
tagger, $\PH$ tagger, and $\cPqt$ tagger, respectively. These taggers make use of the
soft drop mass of AK8 jets, whose distribution after the event selection
is shown in Fig.~\ref{fig:jetmass} for data, a simulated signal for each $\Tp$ decay mode, and the simulated SM
backgrounds. No corrections to the SM backgrounds from the fit to data (explained in Section~\ref{sec:background}) are applied in this figure. 
The selection criteria of the different taggers are:
\begin{itemize}
\item $\ZW$ tagger: AK8 jets are denoted $\ZW$-tagged if their soft drop
  jet mass is in the range $60 < \Msd < 115\GeV$
  and their N-subjettiness ratio
  fulfils $\tau_{21} < 0.5$.
\item $\PH$ tagger: two different $\PH$ taggers are used:
\begin{itemize}
\item[--] $\Hbb$ tagger: AK8 jets are denoted $\Hbb$-tagged if their
  soft drop jet mass is in the range
  $100 < \Msd < 150\GeV$ and two subjet $\cPqb$ tags are
  found. This more stringent selection is used to reduce backgrounds in regions
  with significant background contributions.
\item[--] $\Hb$ tagger: AK8 jets are denoted $\Hb$-tagged if their
  soft drop jet mass is in the range
  $100 < \Msd < 150\GeV$ and exactly one
  subjet $\cPqb$ tag is found. This less stringent selection is used in regions with
  low background contributions.
\end{itemize}
\item $\cPqt$ tagger: AK8 jets are denoted $\cPqt$-tagged if their soft drop jet
  mass is in the range $150 < \Msd <
  220\GeV$ and their N-subjettiness ratio fulfils $\tau_{32} < 0.57$.
\end{itemize}

Events are required to contain at least one $\Hbb$-tagged jet, one
$\Hb$-tagged jet, or one $\ZW$-tagged jet. Because of the overlapping
mass windows in the tagger definitions, an AK8 jet may be tagged by
several taggers. In this case, the priority is given in the following
order to ensure the best signal sensitivity: $\Hbb$, then $\Hb$, and finally $\ZW$,
such that events can be categorised
into three exclusive event categories based on the presence of $\Hbb$-,
$\Hb$-, and $\ZW$-tagged jets.  To maintain sensitivity to both
merged and resolved top quark decays, each of these
three categories is further split into two subcategories, containing
events with and without a $\cPqt$-tagged AK8 jet, respectively. The
resulting six exclusive event categories are listed in the first
column of Table~\ref{tab:categories}, which shows the selection efficiency
for each decay channel of the $\Tp$ in each event category for a signal
with $\MZp=2.5\TeV$ and $\MTp=1.3\TeV$. The selection requirements include all
aforementioned requirements, along with requirements on the reconstructed $\ttbar$
system, detailed in the following section.
The four $\PH$ tag categories feature a higher selection efficiency for
the decay $\Tht$, while the decays $\Tzt$ and $\Twb$ are selected in the
two $\ZW$ tag categories for both the resolved and boosted top quark scenarios.
These decays channels are also selected by the $\PH$ tag categories,
as they are prioritised over the $\ZW$ tag categories.
Signal events with the $\cPqt\PH\cPqt$ final state are reconstructed in the two $\ZW$ tag categories 
if the $\cPqb$ tag criteria for the subjets are not fulfilled, but the $\ZW$ tagging requirements are met.
The selection efficiency is lower in categories requiring a $\cPqt$ tag, since any
top quark produced by this chosen signal will not be significantly boosted,
and therefore will not be efficiently identified by the $\cPqt$ tagger.

\begin{table*}[htb]
\centering
\topcaption{Signal selection efficiency for the three $\Tp$ decay modes in each category for a signal with $\MZp=2.5\TeV$ and $\MTp=1.3\TeV$, taking into account branching fractions $\mathcal{B}(\cPqt\PH\cPqt \to \ljets) = 0.294$, $\mathcal{B}(\cPqt\PZ\cPqt \to \ljets) = 0.317$, and $\mathcal{B}(\cPqt\PW\cPqb \to \ljets) = 0.255$~\cite{PhysRevD.98.030001}, where $\ljets$ is a final state with exactly one electron or muon originating from the decay of one of the top quarks, including electrons and muons from tau lepton decays. The last row of the table shows the total selection efficiency summed over all six categories. 
The efficiencies are shown after all selection requirements, including those on the reconstructed $\ttbar$ system as detailed in Section~\ref{sec:kin_reco}.}
\begin{tabular}{lccc}
Category & {\Tht} [\%] & {\Tzt} [\%] & {\Twb} [\%]\\
\hline
$\Hbb \text{ tag }+ \cPqt \text{ tag }$ & 0.35 & $<$0.1 & $<$0.1 \\
$\Hbb \text{ tag }+ \text{no } \cPqt \text{ tag }$ & 1.7 & 0.15 & $<$0.1 \\
$\Hb \text{ tag }+ \cPqt \text{ tag }$ & 0.93 & 0.12 & $<$0.1 \\
$\Hb \text{ tag }+ \text{no } \cPqt \text{ tag }$ & 5.5 & 1.9 & 0.93 \\
$\ZW \text{ tag }+ \cPqt \text{ tag }$ & 0.33 & 0.15 & $<$0.1 \\
$\ZW \text{ tag }+ \text{no } \cPqt \text{ tag }$ & 2.8 & 7.5 & 5.4 \\ [\cmsTabSkip]
Sum & 11.5 & 11.2 & 6.6 \\
\end{tabular}
\label{tab:categories}
\end{table*}

\subsection{Mass reconstruction}
\label{sec:kin_reco}
The reconstructed $\cPZpr$ mass $\Mrec$ is used as the discriminating observable
between background and signal in this analysis. In addition to the $\ZW$ or Higgs boson,
the signature also requires a $\ttbar$ pair. The reconstruction of a fully resolved $\ttbar$
system is performed by defining top quark \textit{candidates}, built from
the four-momenta of the reconstructed objects.  One candidate is
constructed for the hadronic decay of the (anti)top quark (denoted the hadronic top quark
candidate), and one for the leptonic decay of the (anti)top quark (denoted the
leptonic top quark candidate).  Objects that are used in the
reconstruction are the $\ptvecmiss$, leptons, AK4
jets, and a $\cPqt$-tagged AK8 jet, if present. Only AK4 jets with $\pt > 30\GeV$ and $\abs{\eta} < 2.4$ are
considered. 
The $\cPqb$ tag information of the AK4 jets is not used in the reconstruction of the $\ttbar$ system, 
since it was found that applying it did not improve the assignment of jets to the top quark candidates. 
To ensure that there is no overlap between the two jet collections,
AK4 jets that overlap with the $\ZW$- or $\PH$-tagged jet
within $\Delta R(\text{tagged AK8 jet, AK4 jet}) < 1.2$ are removed from
the event.  If an event has a $\cPqt$-tagged jet, AK4 jets with
$\Delta R(\cPqt\text{-tagged AK8 jet, AK4 jet}) < 1.2$ are removed from the analysis.
Each possible possibility for assigning these objects to the $\ttbar$
system is considered a \textit{hypothesis}.  The best hypothesis is chosen by a
$\chisq$ discriminator that is a measure of the quality of the
reconstruction, combining information from both the hadronic and
leptonic reconstructed top quark candidates. The procedure of building the hypotheses
and calculating $\chisq$ is described in detail below.

The reconstruction of the leptonic top quark candidate requires a neutrino.
Since neutrinos are not measurable in the detector, $\ptvecmiss$ is used to
infer the four-momentum of the neutrino.  It is assumed that
neutrinos are the only source that contributes to $\ptvecmiss$, and thus the $x$ and $y$
components of the neutrino's four-momentum are taken directly from $\ptvecmiss$.
Assuming an on-shell $\PW$ boson, the $z$
component of the neutrino can be calculated by solving the quadratic
equation relating the four-momenta of the $\PW$ and its decay products:
\begin{linenomath}
\begin{equation}
p_{\PW}^2 = M_{\PW}^2 = (p_{\ell} + p_{\nu})^2.
\end{equation}
\end{linenomath}
This quadratic equation can have zero, one, or two real
solutions. In the case of zero real solutions only the real part of the
complex solution is taken.  Candidates are built for each of the
neutrino solutions. In addition to the estimated neutrino momentum, the lepton
momentum is assigned to the leptonic top quark candidate.

If an event has a $\cPqt$-tagged jet, at least one additional AK4 jet is required.
Since the hadronic top quark candidate is already determined, all remaining AK4 jets
in the event are assigned either to the leptonic top quark or are not assigned to a
candidate at all. If an event has no $\cPqt$-tagged jet, at least two additional AK4
jets are required. All AK4 jets in the event are assigned either to the leptonic top
quark candidate, the hadronic top quark candidate, or neither candidate, constructing
all possible candidates.
The four-momenta of the
leptonically and hadronically decaying top quarks are then calculated by
summing the four-momenta of the corresponding
objects (lepton, $\ptmiss$, AK4 jets, and $\cPqt$-tagged AK8 jet, if present).

\begin{figure}[htb]
\centering
\includegraphics[width=0.49\textwidth]{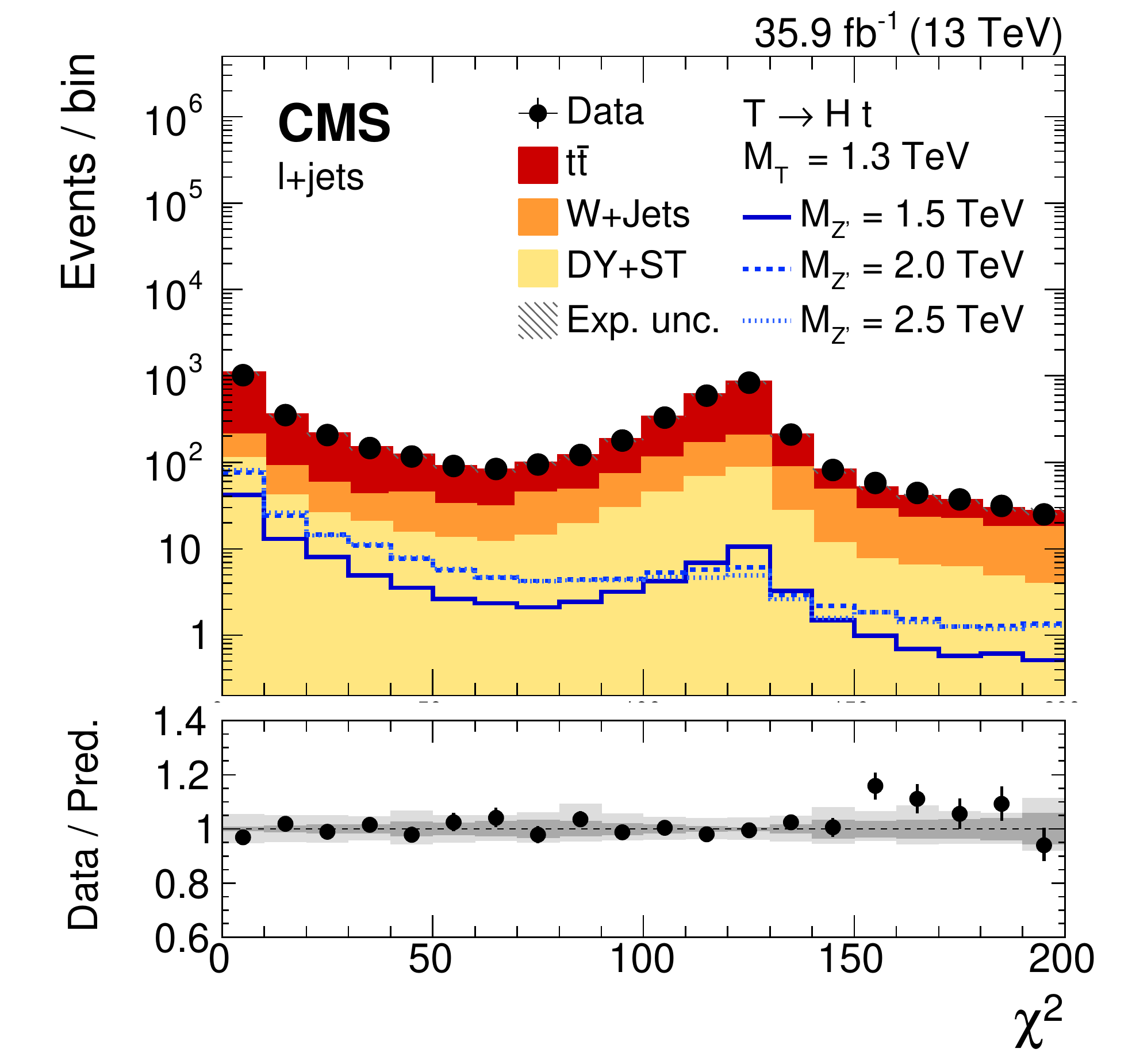}
\caption{Distribution of the smallest $\chisq$ discriminator in each event for the
  combination of both top tag and no top tag categories, after the
  $\ttbar$ reconstruction, combining both lepton channels.
  The simulated SM backgrounds are shown without corrections 
  from the fit to data (explained in Section~\ref{sec:background}).
  The expected signal distribution is shown for various $\MZp$ masses for a fixed mass
  $\MTp=1.3 \TeV$ in the $\Tht$ decay channel, each with a nominal cross section $\sigma(\cPZpr \to \cPqt\Tp)$ of 1\pb.
  The lower panel shows the ratio of data to predicted background. Here the darker grey band indicates the statistical uncertainty, whilst the lighter grey band shows the combined statistical and systematic uncertainty.}
\label{fig:chisq}
\end{figure}

Out of all possible $\ttbar$ hypotheses only one is chosen, based on the smallest value of
$\chisq$, defined as
\begin{linenomath}
\begin{equation}
\chisq =  \left [ \frac{M_{\text{lep}} - \overline{M}_{\text{lep}}}{\sigma_{M_{\text{lep}}}} \right ]^2
            + \left [ \frac{M_{\text{had}} - \overline{M}_{\text{had}}}{\sigma_{M_{\text{had}}}} \right ]^2,
\end{equation}
\end{linenomath}
where $M_{\text{lep}/\text{had}}$ is the invariant mass of the reconstructed
leptonic/hadronic top quark, and $\overline{M}_{\text{lep}/\text{had}}$ and
$\sigma_{M_{\text{lep}/\text{had}}}$ are the average mass and resolution, respectively, of
reconstructed top quark candidates in simulation.
The quantities $\overline{M}_{\text{lep}/\text{had}}$ and
$\sigma_{M_{\text{lep}/\text{had}}}$ are determined from $\ttbar$ simulation
by fitting each of the reconstructed top quark mass distributions with a Gaussian distribution.
The $\cPqb$ quark of the leptonic top quark decay from simulation is required to match the
assigned reconstructed jet within $\Delta R (\cPqb,\, \jet) <0.4$,
whilst the jets used to reconstruct the hadronic top quark are required to
match with the quarks from the hadronic top quark
decay from simulation within $\Delta R (\Pq,\, \jet) <0.4$.
The distribution of the smallest $\chisq$ discriminator in each event is shown in Fig.~\ref{fig:chisq}.
The $\chisq$ discriminator tends to zero for well-reconstructed $\ttbar$ systems,
and to higher values for poor quality reconstructions and background events.
In events where only one top quark is well-reconstructed,
the $\chisq$ peaks at values of $\chisq \approx 120$.
From optimisation studies, it was found that requiring events in the
signal region to have $\chisq < 50$ ensured the best sensitivity.

Finally, $\Mrec$ is calculated by
summing all four momenta of the chosen $\ttbar$ hypothesis along with
that of the tagged $\ZW$ or Higgs boson.

The impact of each stage of the selection requirements on the signal selection
efficiency for each decay channel of the $\Tp$ is shown in Table~\ref{tab:cutflow}
for a signal with $\MZp=2.5\TeV$ and $\MTp=1.3\TeV$.
The lower efficiency in the $\Twb$ decay channel is primarily due to the requirement placed on $\chisq$,
since there is only one $\cPqt$ quark emitted in the decay chain for this channel.

\begin{table*}[htb]
\centering
\topcaption{Signal selection efficiency after each step in the selection requirements for a signal with $\MZp=2.5\TeV$ and $\MTp=1.3\TeV$, taking into account branching fractions $\mathcal{B}(\cPqt\PH\cPqt \to \ljets) = 0.294$, $\mathcal{B}(\cPqt\PZ\cPqt \to \ljets) = 0.317$, and $\mathcal{B}(\cPqt\PW\cPqb \to \ljets) = 0.255$~\cite{PhysRevD.98.030001}, where $\ljets$ is a final state with exactly one electron or muon originating from the decay of one of the top quarks, including electrons and muons from tau lepton decays.}
\begin{tabular}{lcccccc}
\multirow{2}{*}{Selection requirement} & \multicolumn{2}{c}{{\Tht} [\%]} & \multicolumn{2}{c}{{\Tzt} [\%]} & \multicolumn{2}{c}{{\Twb} [\%]}\\
& $\mu$ & $e$ & $\mu$ & $e$ & $\mu$ & $e$ \\
\hline
Trigger and exactly 1 muon (electron) &\multirow{2}{*}{57} &  \multirow{2}{*}{14} &\multirow{2}{*}{45} &\multirow{2}{*}{14}& \multirow{2}{*}{66}& \multirow{2}{*}{30}\\
with $\pt > 50~(125)\GeV$ and $\abs{\eta} < 2.4$ &   &   & & & &  \\[0.1cm]
$\geq$1 AK8 jet with $\pt > 250 \GeV$  &  52 &13&  39&13 &  56&25\\[0.1cm]
$\ptrel(\ell,\, \jet) > 40\GeV$  or  $\Delta R(\ell,\, \jet)> 0.4$  &  25 &  9.6&21 &9.3&  34&21\\[0.2cm]
$\geq$1 $\PH$ or $\ZW$ tag  &  14 &5.3&  13 &5.6&  15&8.2\\[0.1cm]
$\geq$1 AK4 jet (top tag cat.) & \multirow{2}{*}{13} & \multirow{2}{*}{5.0}&\multirow{2}{*}{12} &\multirow{2}{*}{5.0}& \multirow{2}{*}{12} & \multirow{2}{*}{6.2}\\
or 2 AK4 jets (no top tag cat.)   &   & & & &   &  \\[0.1cm]
$\chisq < 50 $   &  8.5 &3.0 &  8.2 &3.0 &  4.6 & 2.0\\
\end{tabular}
\label{tab:cutflow}
\end{table*}

\section{Background estimation}
\label{sec:background}

A multistep procedure is performed to ensure that the Monte Carlo (MC) simulation
(Section~\ref{sec:datasets}), used to estimate the backgrounds, accurately describes the data.

We apply scale factors to the simulation to
account for the measured differences between simulation and data
in the mistag rates and tagging efficiencies for the $\ZW$, $\PH$, and $\cPqt$ taggers.
The mistag rates are measured both in data and in simulation using a QCD
multijet-enriched region, while the tagging efficiencies are measured in a
\ttbar-enriched region. Finally, in the statistical analysis simulations are
constrained using control regions in data.
These are fit simultaneously with the signal regions, constraining the
normalizations and shapes of the background distributions while efficiently
searching for a signal.

The mistag rate is determined from a QCD multijet-enriched data sample where the contribution from real \PZ, $\PW$, or Higgs bosons is negligible.
The mistag rate is defined as the number of AK8 jets after the tagger is applied, divided by the number of AK8 jets before the tagger is applied.
The data are selected with a trigger requiring the scalar \pt sum of the jets in the event, defined as \HT, to be $\HT>900\GeV$.
The selected data events are then required to have $\HT > 1000\GeV$ to ensure that events are selected in a region of phase space where the trigger is fully efficient.
The AK8 jets must have $\abs{\eta} < 2.5$, $\pt > 200\GeV$, and $\Msd > 30\GeV$.
The uncertainties in the mistag rate scale factors receive contributions from statistical uncertainties,
and from effects associated with differences in the quark and gluon compositions and in the
kinematic distributions between the QCD multijet and $\ZW$+jets samples.
The mistag rate for the $\ZW$ tagger is measured both in data and in simulation, resulting in a scale factor of $1.05 \pm 0.08$ that is applied to simulation in the signal regions.
The scale factor is only applied to jets that are $\ZW$ tagged and are not matched to a $\ZW$ boson at generator level, where a jet is considered matched if both quarks of the hadronic boson decay are within $\Delta R (\Pq,\, \text{tagged AK8 jet}) < 0.8$.
The mistag rate scale factors for the $\Hbb$ and $\Hb$ taggers are $1.15 \pm 0.18$ and $1.22 \pm 0.05$, respectively, and are applied to all MC samples (except \ttbar) that do not contain real Higgs bosons.
Since the fraction of jets initiated by a $\cPqb$ quark is significantly higher in the \ttbar background than in the \Wjets and QCD multijet backgrounds,
a dedicated scale factor is calculated for the mistag rate of the $\Hbb$ and $\Hb$ taggers in a sample of \ttbar events with back-to-back topology in the $\ljets$ final state.
These mistag rate scale factors are measured to be $1.01 \pm 0.18$ and $0.99 \pm 0.03$ for the $\Hbb$ and $\Hb$ taggers, respectively.
The mistag rate scale factor for the $\cPqt$ tagger is measured to be $0.95 \pm 0.02$.
The scale factor is only applied to jets that are $\cPqt$ tagged and do not match to a top quark at generator level, where a match requires the three quarks from the hadronic top quark decay to have $\Delta R (\Pq,\, \text{tagged AK8 jet}) < 0.8$.

The efficiency scale factor for the $\cPqt$ tagger is measured using the procedure described in Ref.~\cite{CMS-PAS-JME-16-003} and is found to be $1.06^{+0.07}_{-0.04}$.

The efficiency of the $\ZW$ tagger is measured in a $\ttbar$-enriched region with a back-to-back topology.
One of the top quarks is required to decay leptonically into a $\cPqb$ quark and a $\PW$ boson, with the $\PW$ boson decaying into a lepton and a neutrino, whilst the other top quark decays hadronically, resulting in a $\cPqb$ quark reconstructed as an AK4 jet and a $\PW$ boson reconstructed as an AK8 jet.
This AK8 jet is used to measure the efficiency of the $\ZW$ tagger.
The sample is selected following the same procedure as in Section~\ref{sec:preselection},
additionally requiring at least two AK4 jets, where each jet must have $\pt > 30\GeV$, $\abs{\eta} < 2.4$,
must pass the medium working point of the $\cPqb$ tagging discriminator, and not overlap with the AK8 jet.
It is required that the angular separation $\Delta R$ between the leptonic top quark and the hadronic top quark is greater than $\pi/2$ in order to reconstruct $\PW$ bosons well-separated from nearby $\cPqb$ quark jets.
The $\ZW$ tagger efficiency scale factor is then estimated with a procedure similar to that used for the $\cPqt$ tagger, and is found to be $0.91 \pm 0.08$ for events in the signal region.
The systematic uncertainty for the dependence of the scale factor on the choice of the fit model used
to extract the boosted $\PW$ contribution from the combinatorial $\ttbar$ background is estimated to be 1\%.
A direct measurement of the tagger efficiency scale factor using data is only possible
for jet $\pt \lesssim 200\GeV$.
For larger jet \pt, the difference between \ttbar samples simulated with two different shower
and hadronization models (\PYTHIA 8 and \HERWIGpp~2.7.1~\cite{Bahr:2008pv}) contributes an
additional uncertainty to the \pt dependence of the scale factor parameterised as $4.1\% \times \ln(\pt/200{\GeV})$.

For the $\Hbb$ and $\Hb$ taggers, the efficiency scale factor of the $\ZW$ tagger is used,
taking into account its associated uncertainty. Since the $\PH$ taggers do not utilise a requirement on $\tau_{21}$, they are not assigned a corresponding \pt-dependent uncertainty. In addition, two further uncertainties are considered.
Firstly, the uncertainty in the extrapolation of the scale factor from the $\ZW$ selection to
the $\Hbb$ and $\Hb$ selections is estimated from the difference between the two shower and
hadronization models (\PYTHIA 8 and \HERWIGpp).
Secondly, uncertainties in the subjet $\cPqb$ tagging efficiencies are also included,
as described in Ref.~\cite{Sirunyan:2017ezt}.

The main background processes that contribute to this analysis are $\ttbar$ and $\Wjets$.
Two control regions, each chosen to enhance a background process, are used to constrain the production rate of these processes and reduce potential mismodelling of the event kinematic variables.
The control regions are also used to verify the agreement of the simulation with data.
They are based on the selection described in Section~\ref{sec:preselection}.
In addition to this selection, we require $\chisq < 50$, and the mass of the $\Hbb$-/$\Hb$-/$\ZW$-tagged
AK8 jet to be either $\Msd < 60\GeV$ or $\Msd > 150\GeV$.
This last requirement ensures events in the control regions are not also found in the signal regions.
The first control region is designed to be enriched in $\ttbar$ events, and is obtained by requiring at least one additional $\cPqb$-tagged AK4 jet.
The second control region is designed to be enriched in $\Wjets$ events, and is obtained by requiring no additional $\cPqb$-tagged AK4 jets.

\begin{figure*}[htb]
\centering
\includegraphics[width=0.39\textwidth]{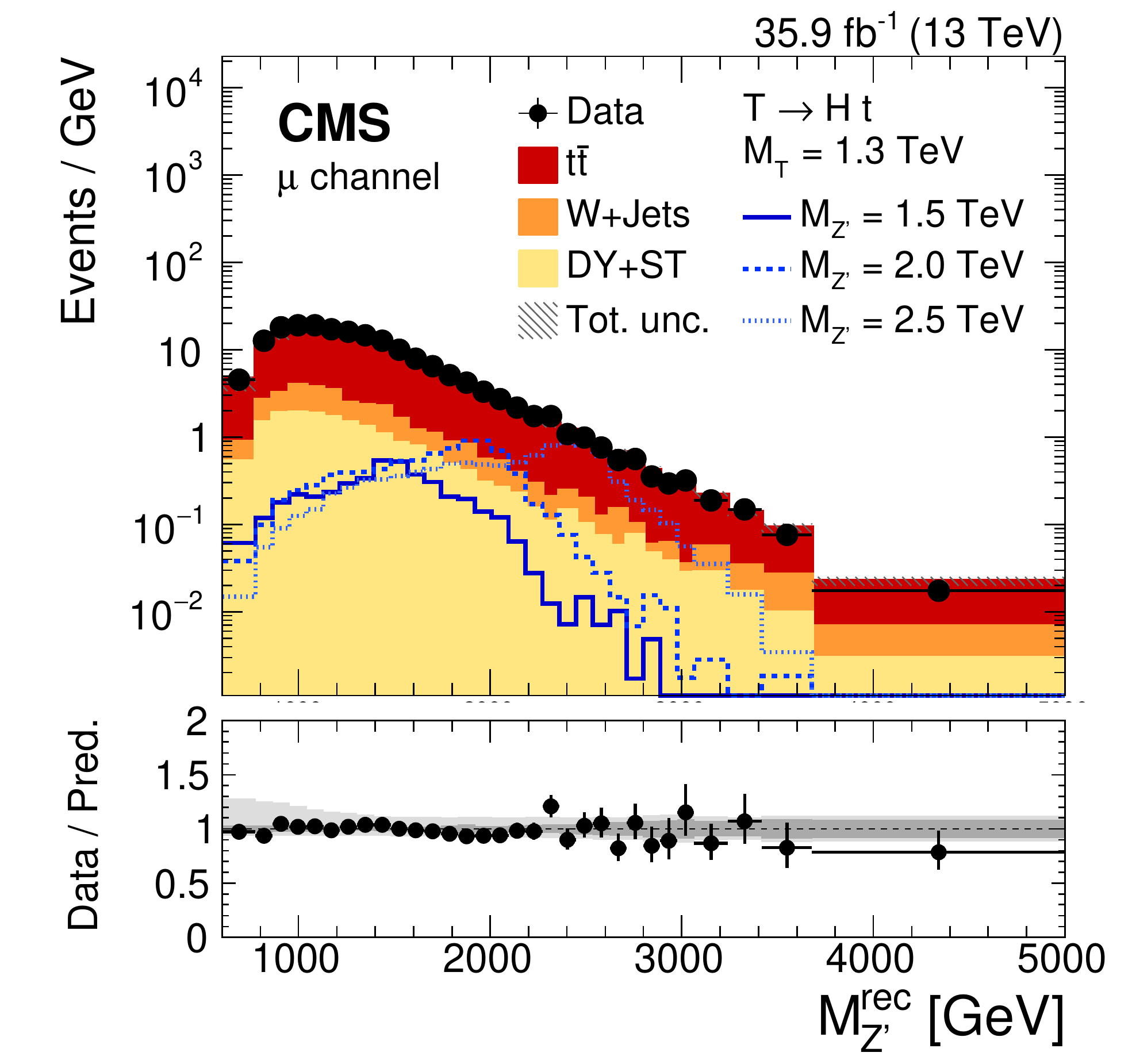}
\includegraphics[width=0.39\textwidth]{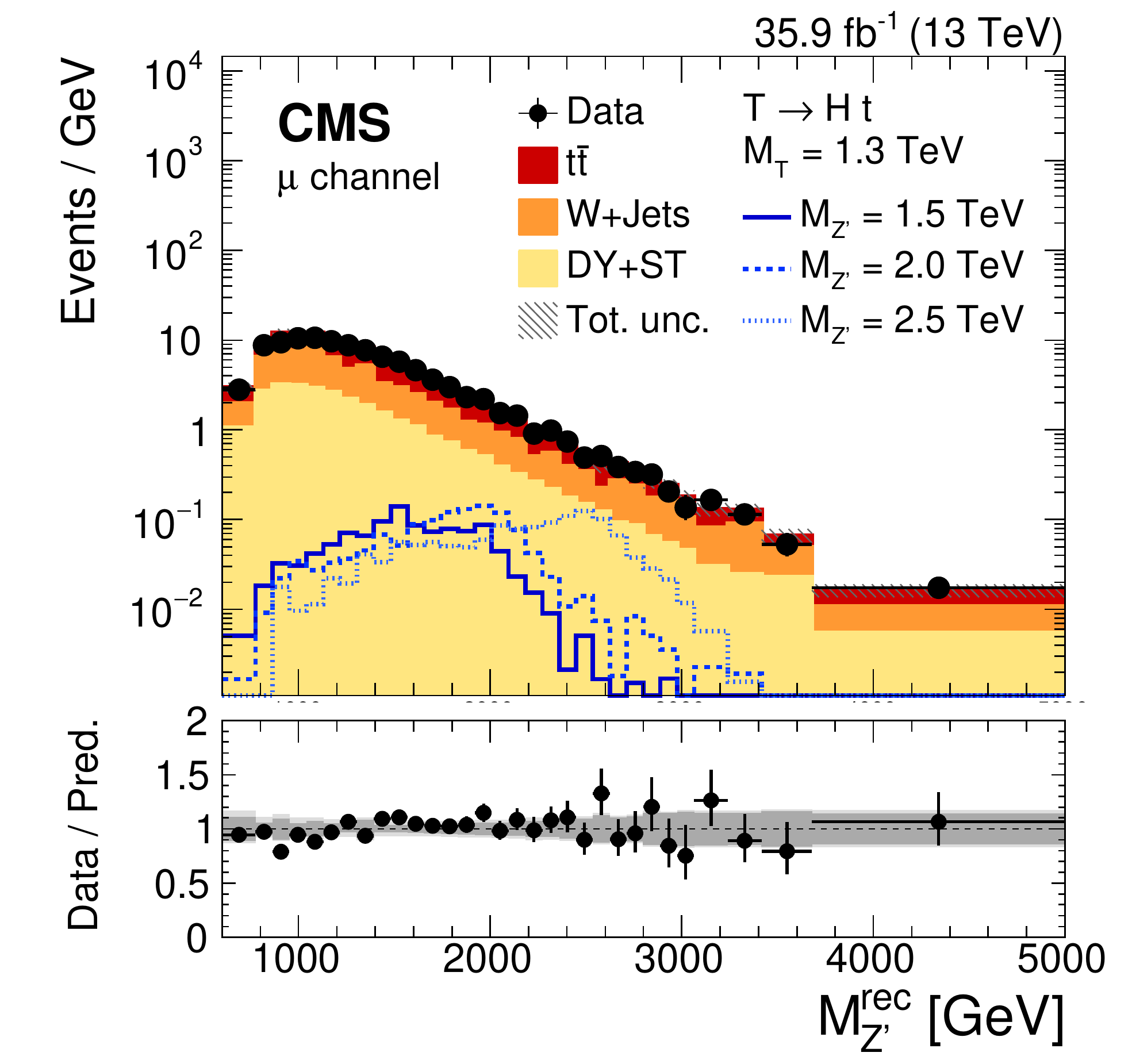} \\
\includegraphics[width=0.39\textwidth]{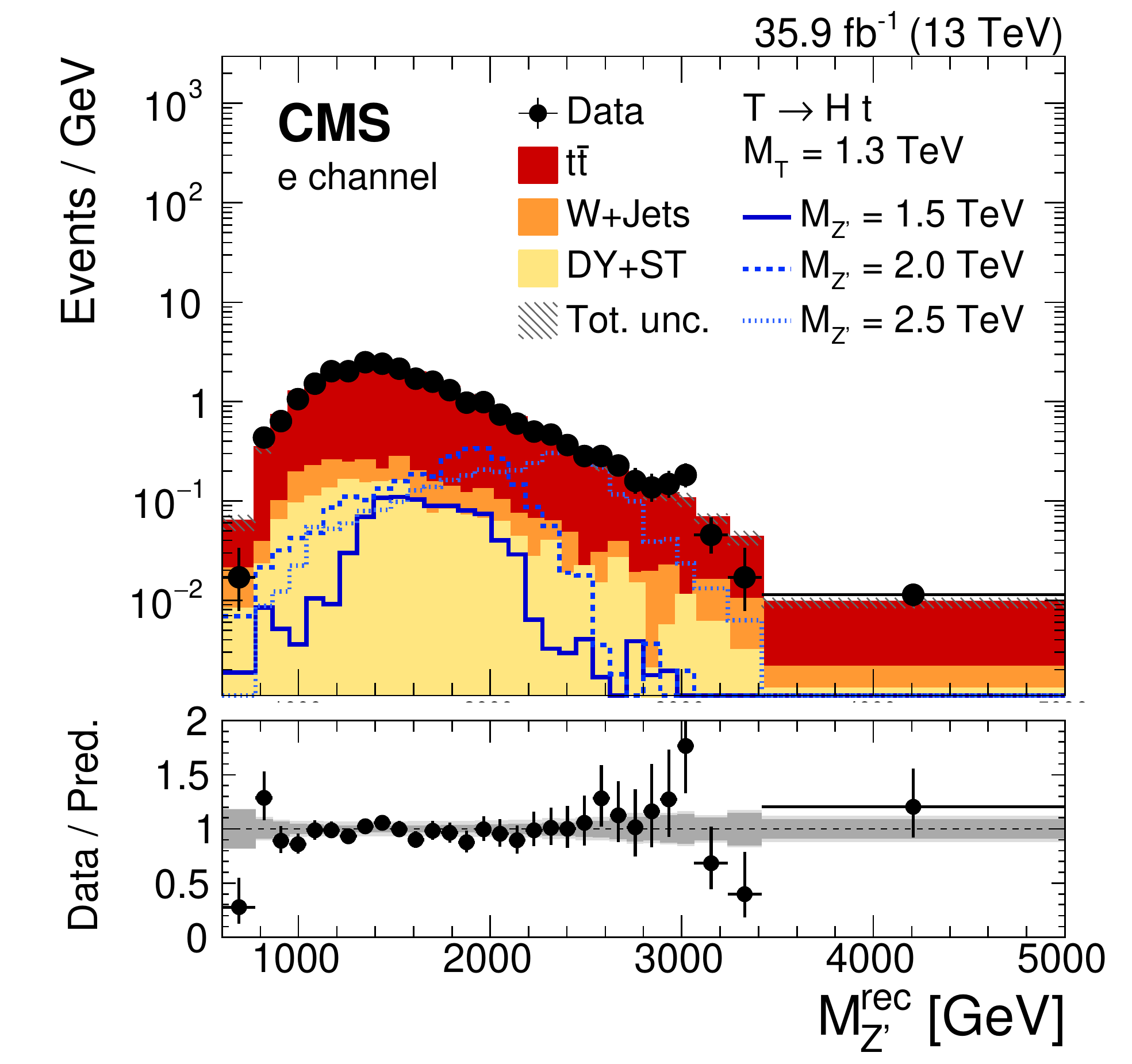}
\includegraphics[width=0.39\textwidth]{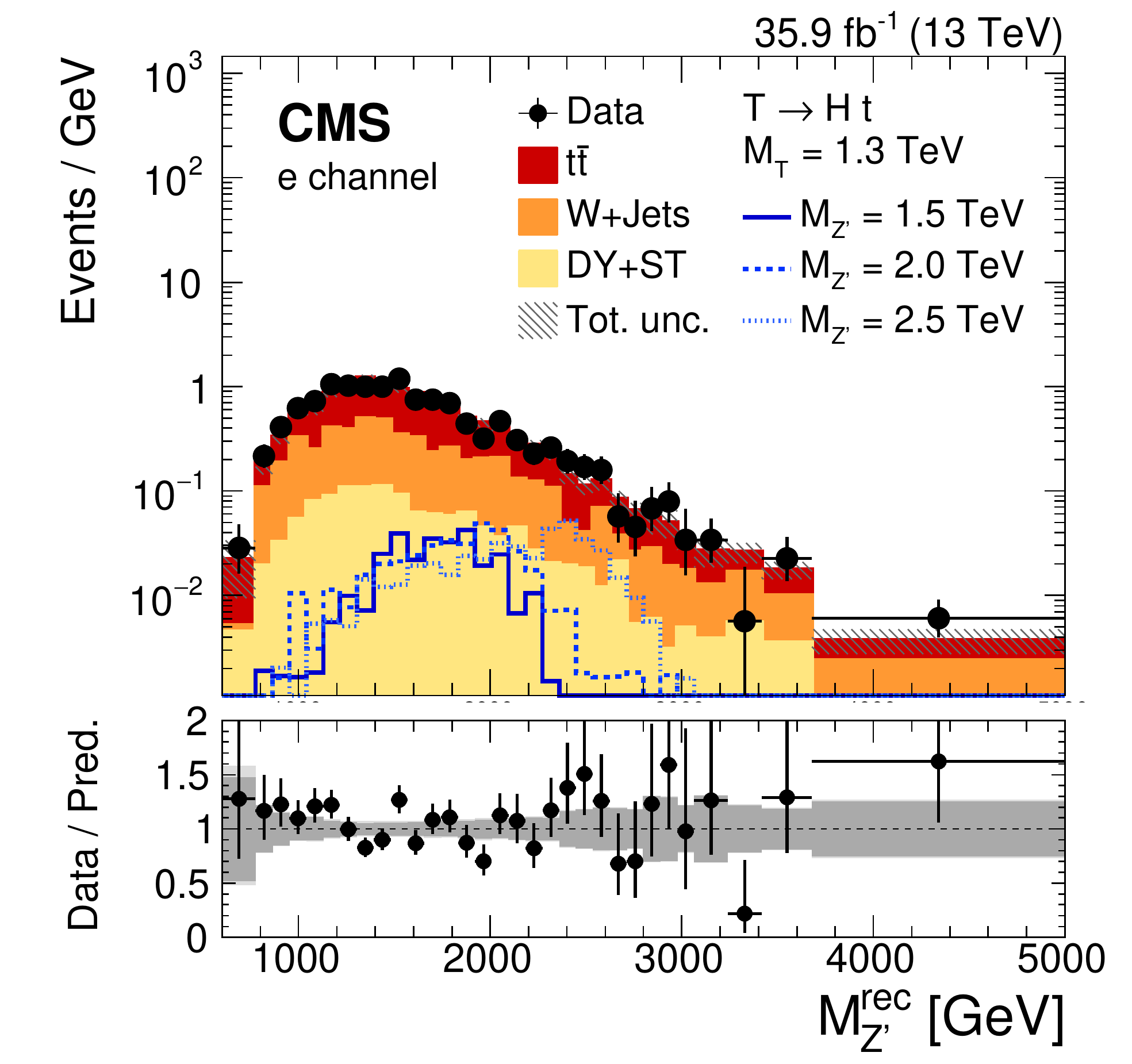}

\caption{Distribution of the reconstructed $\cPZpr$ boson mass in the
  {\mujets} channel (upper row) and {\ejets} channel (lower row) for the $\ttbar$-enriched control region (left)
  and for the $\Wjets$-enriched region (right). The expected signal distribution is shown for various $\MZp$ masses for a fixed mass
  $\MTp=1.3\TeV$ in the $\Tht$ decay channel, each with a nominal cross section $\sigma(\cPZpr \to \cPqt\Tp)$ of 1\pb.
  The lower panel shows the ratio of data to predicted background. Here the darker grey band indicates the statistical uncertainty, whilst the lighter grey band shows the combined statistical and systematic uncertainty.}
\label{fig:muon_sideband}
\end{figure*}

Figure~\ref{fig:muon_sideband} shows the $\Mrec$ distribution in the control regions for the muon and electron channels, after fitting the $\ttbar$ and $\Wjets$ backgrounds simultaneously in both control and signal regions.
It can be seen that there is good agreement between data and simulation.
Similar agreement is found in kinematic distributions of the objects used to
reconstruct the $\cPZpr$ resonance mass.
Both control regions are included in the maximum-likelihood based fit described in Section~\ref{sec:results}.
The fit estimates the size of a possible signal, whilst simultaneously constraining the
background simulation normalizations using the data in the control regions.

\section{Systematic uncertainties}
\label{sec:systematics}

Systematic uncertainties can affect both the normalization and the shape of the $\Mrec$ distributions.
The uncertainties considered in this analysis are explained in the following and listed in Table~\ref{tab:uncertainties}.

For each AK8 jet, the combined statistical and systematic uncertainty in the $\ZW$, $\PH_{2\cPqb/1\cPqb}$ and $\cPqt$ tagger efficiency scale factors and misidentification rate scale factors for the $\ZW$, $\PH_{2\cPqb/1\cPqb}$, and $\cPqt$ taggers, are propagated to variations of signal and background distributions.

Uncertainties in the jet energy scale~\cite{Khachatryan:2016kdb} have been measured as a function of \pt and $\eta$.
The jet energy scale is varied within ${\pm}1$ standard deviation for AK8 and AK4 jets simultaneously.
The jet energy scale uncertainty is also propagated to $\ptmiss$.

The uncertainty in the jet energy resolution has been measured in different $\eta$ bins~\cite{Khachatryan:2016kdb}.
This uncertainty is applied to AK4 and AK8 jets simultaneously, assessing the impact of varying their resolutions by ${\pm}1$ standard deviation. The variation is also propagated to $\ptmiss$.

The $\cPqb$ tagging efficiencies are measured in a sample enriched with heavy-flavour jets, whilst the probability to tag a jet originating from a different flavour as a $\cPqb$ quark jet (a mistag) is measured in a sample enriched with light-flavour jets. These are applied to jets in signal and background events~\cite{Sirunyan:2017ezt}.
The uncertainties in these measurements are propagated to variations of signal and background normalizations and shapes.

Data-to-simulation scale factors for muon and electron identification and trigger efficiencies
are applied as a function of \pt and $\eta$.
The effect of varying each scale factor by ${\pm}1$ standard deviation
is studied to estimate uncertainties in the normalizations and shapes of the signal and
background distributions.

The uncertainty in the integrated luminosity of the 2016 data set is 2.5\%~\cite{CMS-PAS-LUM-17-001}.
The effect of pileup is studied by comparing simulated samples
where the distribution of pileup interactions is varied according to its uncertainty.

The uncertainties in the factorisation and renormalization scales $\muf$ and $\mur$ are taken into account for $\Wjets$, $\Zjets$, $\ttbar$, and single top quark backgrounds, as well as for the signal.
The uncertainty related to the choice of $\muf$ and $\mur$ scales is evaluated following the proposal in Refs.~\cite{Cacciari:2003fi,Catani:2003zt} by varying the default choice of scales by the following six combinations of factors,
$(\muf, \mur) \times (1/2, 1/2)$, $(1/2, 1)$, $(1, 1/2)$, $(2, 2)$, $(2, 1)$, and $(1, 2)$.
The maximum and minimum of the six variations are computed for each bin of the $\Mrec$ distribution, producing an uncertainty ``envelope'' that affects both normalization and shape.

For samples generated at LO and NLO, uncertainties based on the NNPDF 3.0 PDF sets~\cite{Ball:2014uwa} and PDF4LHC15 (NLO 100)~\cite{Butterworth:2015oua,Carrazza:2015hva}, respectively, have been evaluated using the PDF4LHC procedure~\cite{Butterworth:2015oua}, where the root-mean-square of 100 pseudo-experiments provided by the PDF sets represent the uncertainty envelope.
For the $\Wjets$, $\Zjets$, $\ttbar$, and single top quark background processes, the full uncertainty in normalization and shape due to the variations in cross section is evaluated.
For signal samples, only the uncertainty in normalization and shape due to the variations in event selection and reconstruction efficiency is taken into account, and overall uncertainties in the inclusive cross section due to PDF variations are only displayed as error bands on benchmark theory model predictions.

\begin{table*}[htb]
\centering
\topcaption{List of systematic uncertainties considered in the statistical analysis, with the size of their impact, the type(s) of effect they have, and the categories they affect. The impact size of each uncertainty is based on a signal sample with $\MZp = 1.5\TeV$ and $\MTp = 1.3\TeV$. All uncertainties affect the
normalizations of the $\Mrec$ distributions. The ones also affecting the shapes are indicated
by a tick mark. Uncertainties that affect control regions are denoted by CR,
whilst those that affect signal regions are denoted by SR.}
\begin{tabular}{lclc}
Source & Uncertainty [\%] & Shape & Categories \\
\hline
$\ZW$ tagging efficiency & $8 \oplus 4.1 \times \ln(\pt/200{\GeV})$&  & $\ZW$ tag\\
$\ZW$ mistag rate & $\pm 5.6\text{--}7.9$ &  \checkmark &  $\ZW$ tag\\
$\Hbb/\Hb$ tagging efficiency & 9 &  & $\Hbb/\Hb$ tag\\
$\Hbb$ mistag rate & $\pm  14\text{--}18 $ &  \checkmark & $\Hbb$ tag\\
$\Hb$ mistag rate & $\pm 3.2\text{--}4.6 $&  \checkmark & $\Hb$ tag\\
$\Hbb$ mistag rate (only $\ttbar$) & 18 &  & $\Hbb$ tag\\
$\Hb$ mistag rate (only $\ttbar$) & 3 &  & $\Hb$ tag\\
$\cPqt$ tagging efficiency & ${+7}/{-4}$ &  & top tag\\
$\cPqt$ mistag rate & 1.8 & \checkmark & top tag\\
Jet energy scale & $\pm 0.1\text{--}5.5$&   \checkmark & CR+SR\\
Jet energy resolution & ${<}0.01$&   \checkmark & CR+SR\\
$\cPqb$ tagging AK4 & $\pm 1.8\text{--}3.0 $ &  \checkmark & CR \\
$\cPqb$ tagging AK8 & $\pm 2.7\text{--}7.3 $ &  \checkmark &$\Hbb/\Hb$ tag \\
Muon ID & $\pm 0.1\text{--}2.6 $&  \checkmark & CR+SR\\
Muon trigger & $\pm 0.4\text{--}2.2  $&  \checkmark & CR+SR\\
Muon tracker & $ \pm 0.5\text{--}1.8 $&  \checkmark & CR+SR\\
Electron ID & $ \pm 0.3\text{--}3.1 $&  \checkmark & CR+SR\\
Electron trigger &$\pm 0.4\text{--}0.5  $ &  \checkmark & CR+SR\\
Electron reconstruction &$\pm 0.1\text{--}3.0 $ &  \checkmark & CR+SR\\
Luminosity & 2.5 &  & CR+SR\\
Pileup reweighting & $\pm 0.1\text{--}3.3 $&  \checkmark & CR+SR \\ [\cmsTabSkip]
$\muf$ and $\mur$ scales & 6 variations &  \checkmark & CR+SR\\
PDF & 100 samples & \checkmark & CR+SR\\
\end{tabular}
\label{tab:uncertainties}
\end{table*}

\section{Results}
\label{sec:results}

\begin{figure*}[htbp!]
\centering
\includegraphics[width=0.39\textwidth]{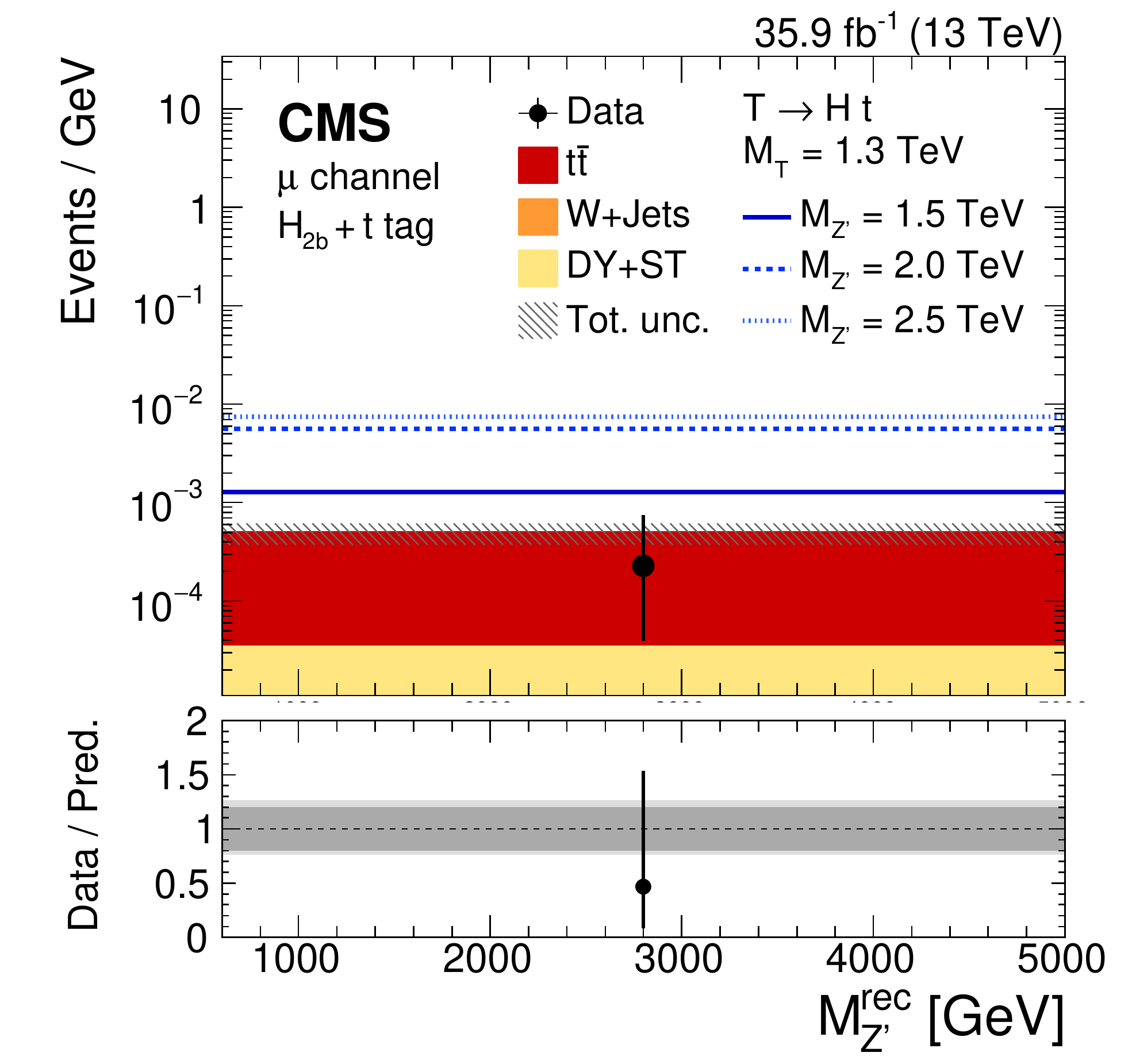}
\includegraphics[width=0.39\textwidth]{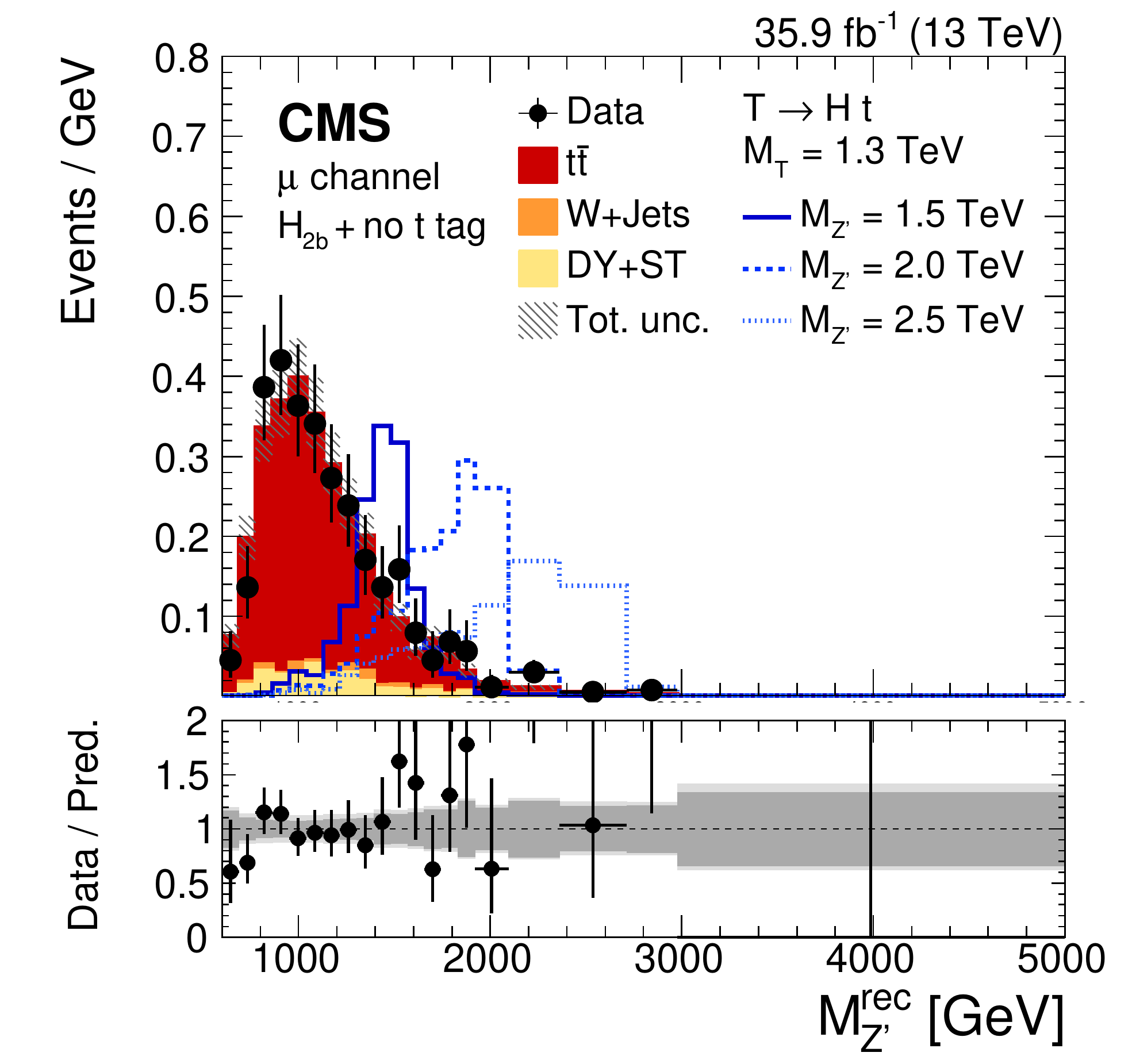} \\
\includegraphics[width=0.39\textwidth]{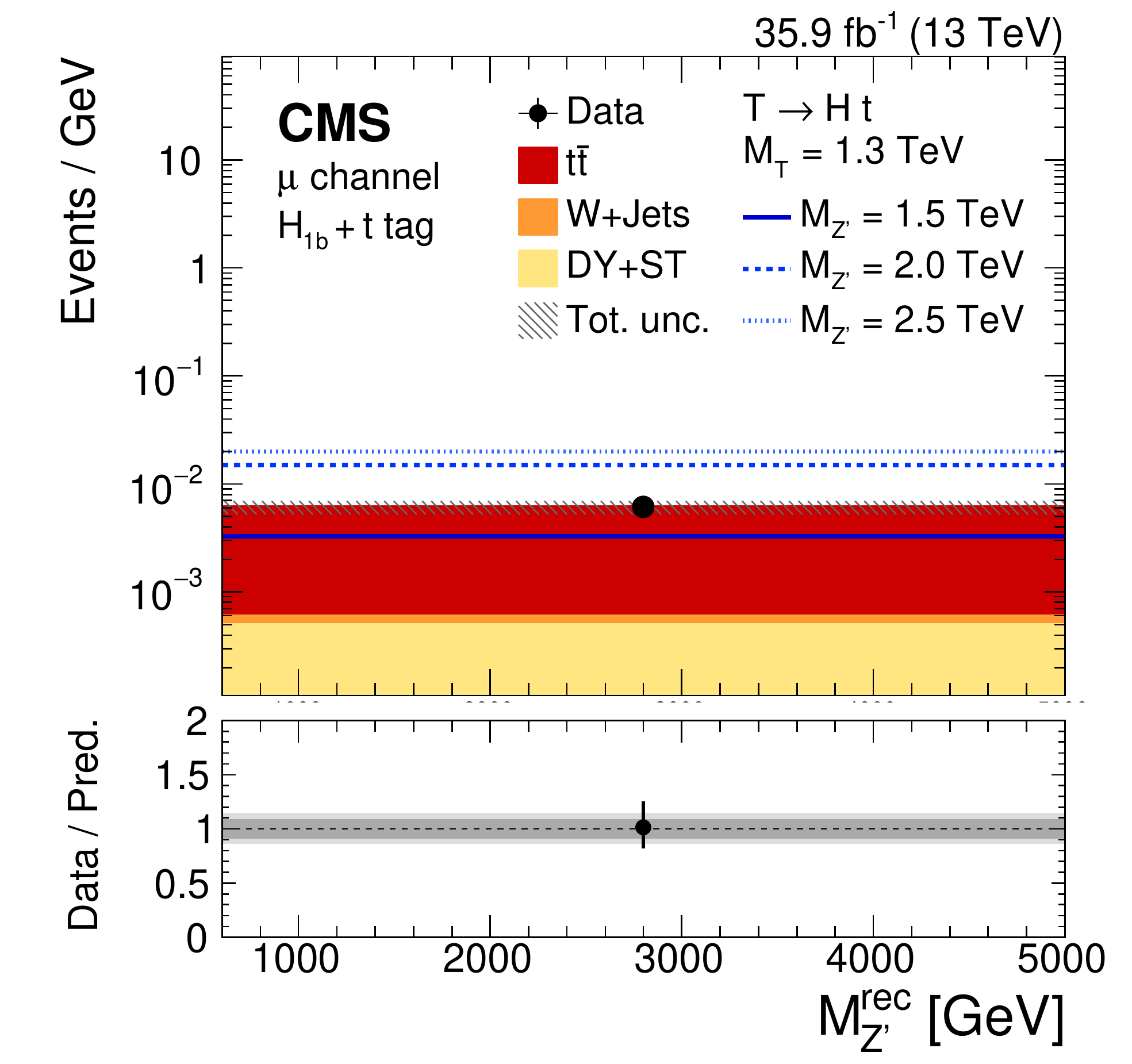}
\includegraphics[width=0.39\textwidth]{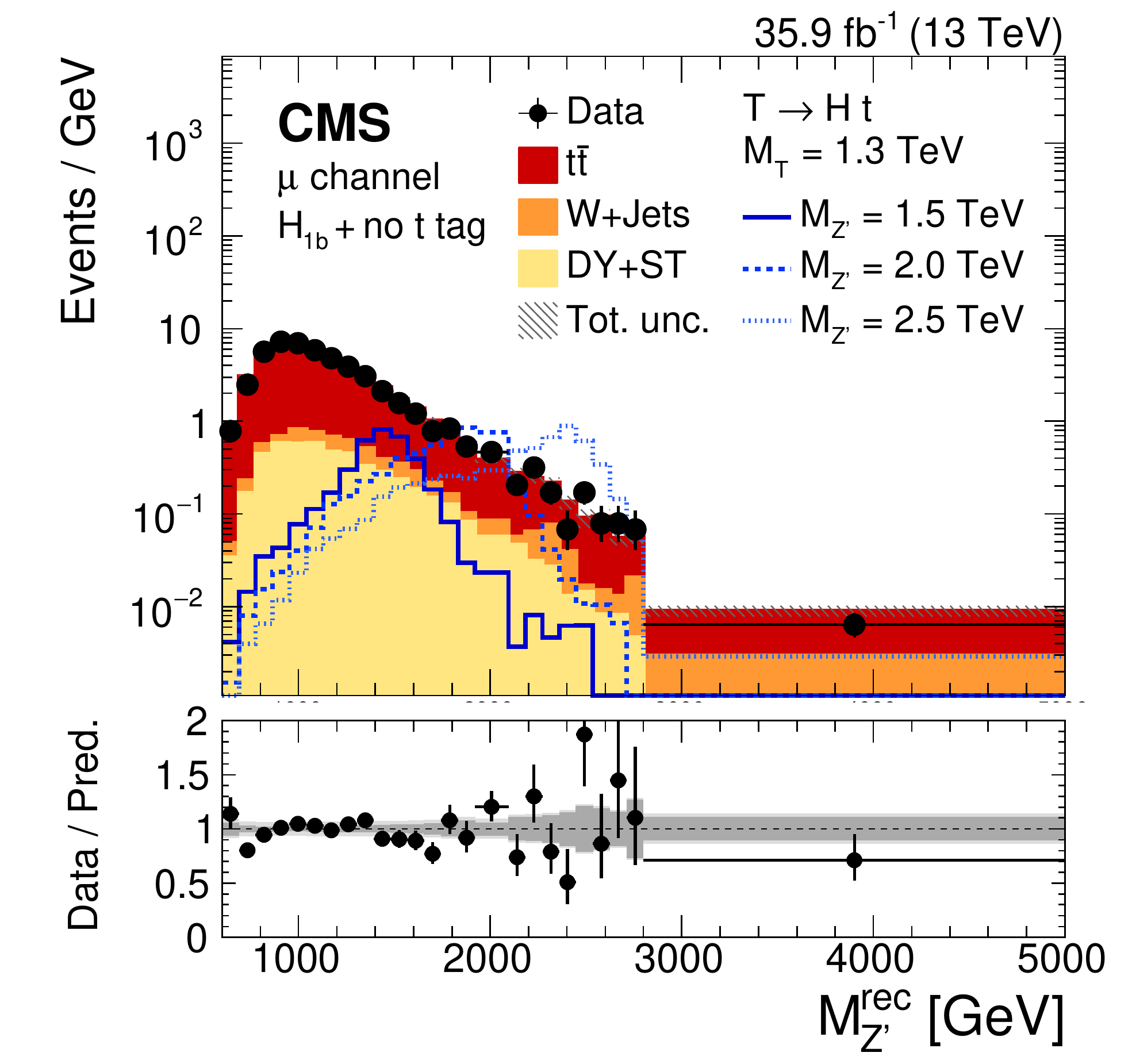} \\
\includegraphics[width=0.39\textwidth]{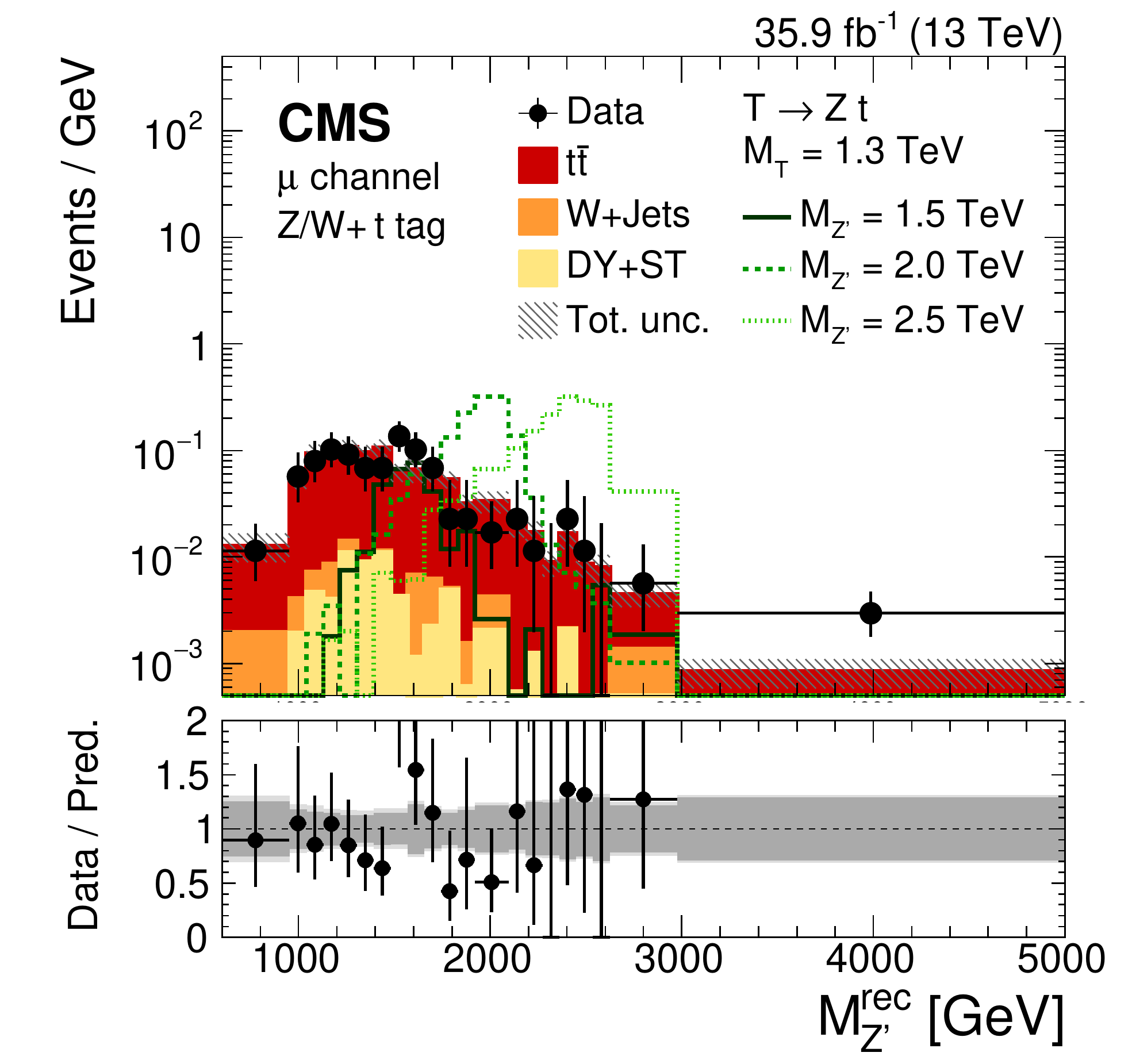}
\includegraphics[width=0.39\textwidth]{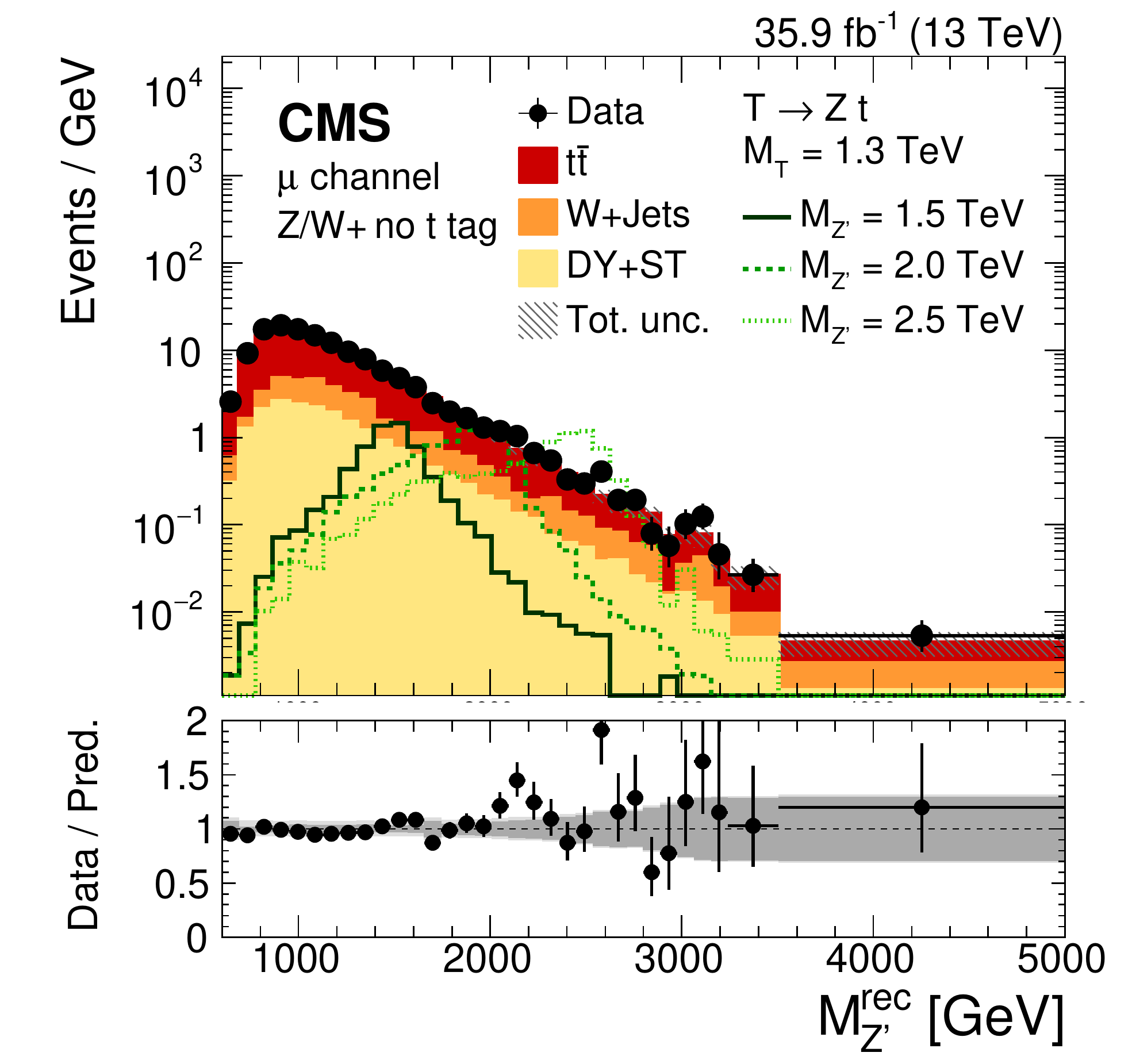}
\caption{Distribution of the reconstructed $\cPZpr$ resonance mass after the
  full selection in the {\mujets} channel for the data, the expected SM
  background, and for the signal with different $\cPZpr$ masses for a fixed
  $\Tp$ mass of 1.3\TeV. In the left (right) column the results in the
  top tag (no top tag) category are shown. Different rows display the
  distributions of events accepted by different taggers as well as the
  signal for the respective $\Tp$ decays: $\Hbb$ tagger and
  $\Tht$ decay (upper), $\Hb$ tagger and $\Tht$ decay (middle), and $\ZW$ tagger and $\Tzt$ decay (lower).
  The signal histograms correspond to a nominal cross section $\sigma(\cPZpr \to \cPqt\Tp)$ of 1\pb.
  The lower panel shows the ratio of data to predicted background. Here the darker grey band indicates the statistical uncertainty, whilst the lighter grey band shows the combined statistical and systematic uncertainty.}
\label{fig:final_muon}
\end{figure*}

\begin{figure*}[htbp!]
\centering
\includegraphics[width=0.39\textwidth]{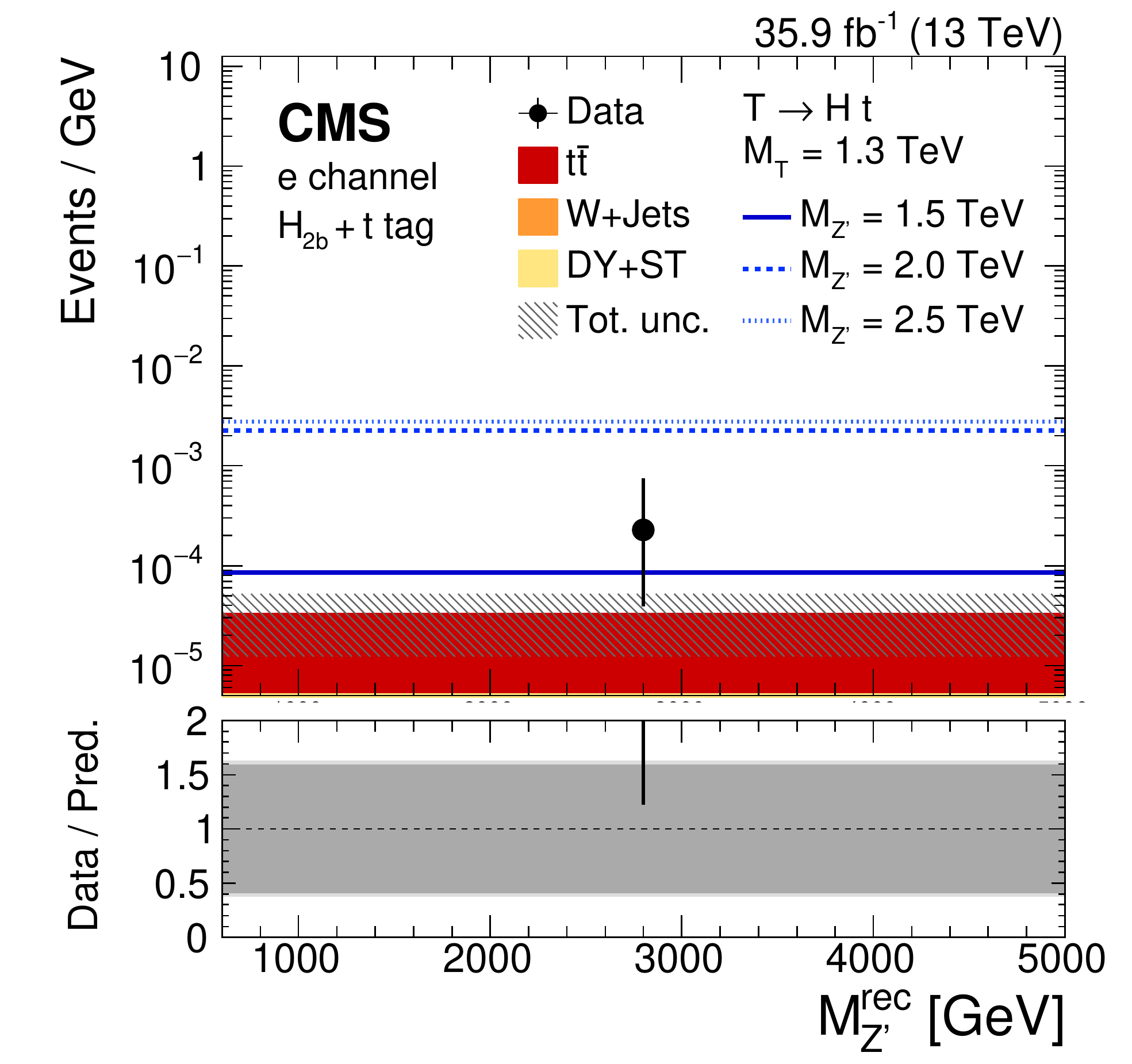}
\includegraphics[width=0.39\textwidth]{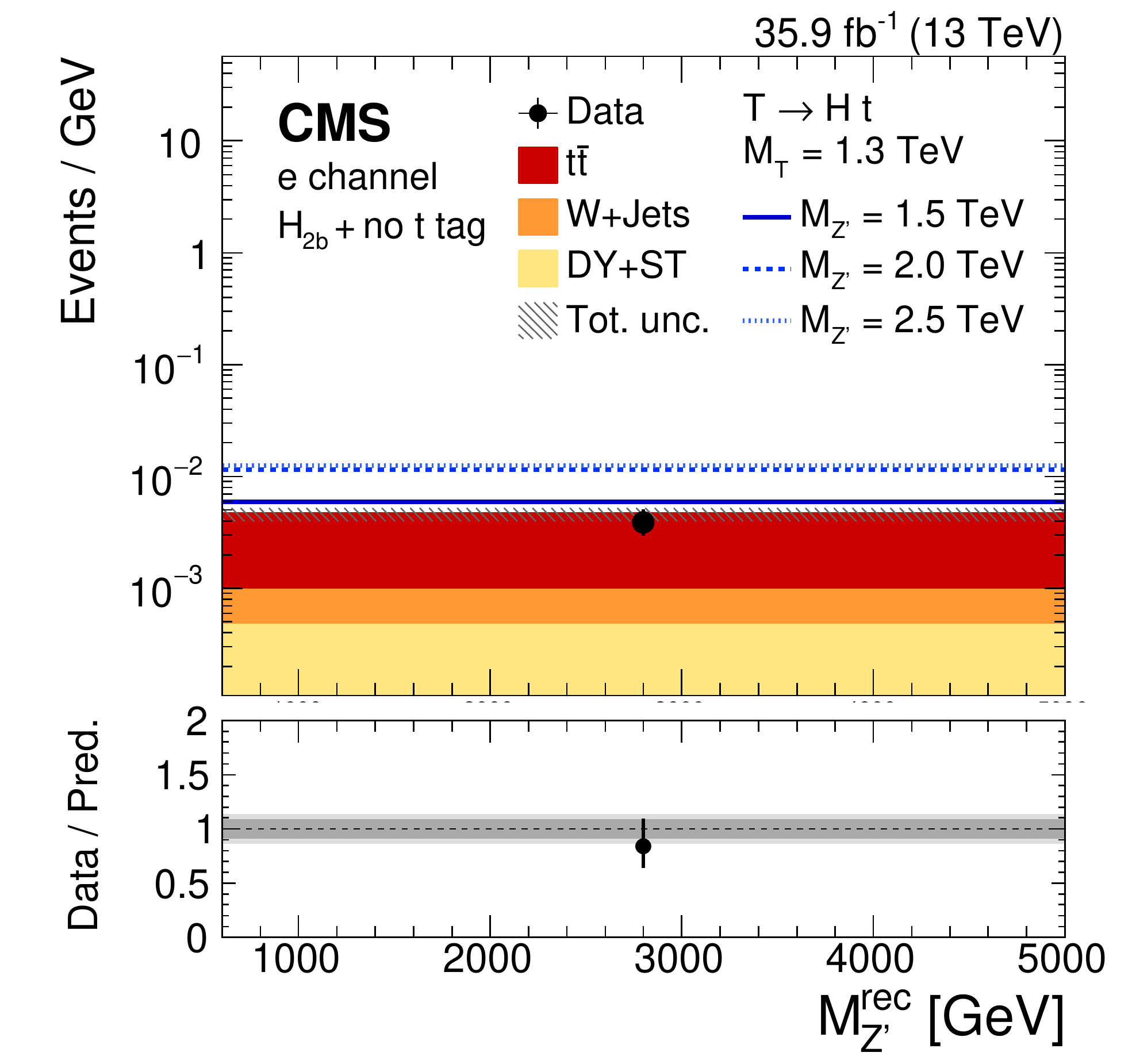} \\
\includegraphics[width=0.39\textwidth]{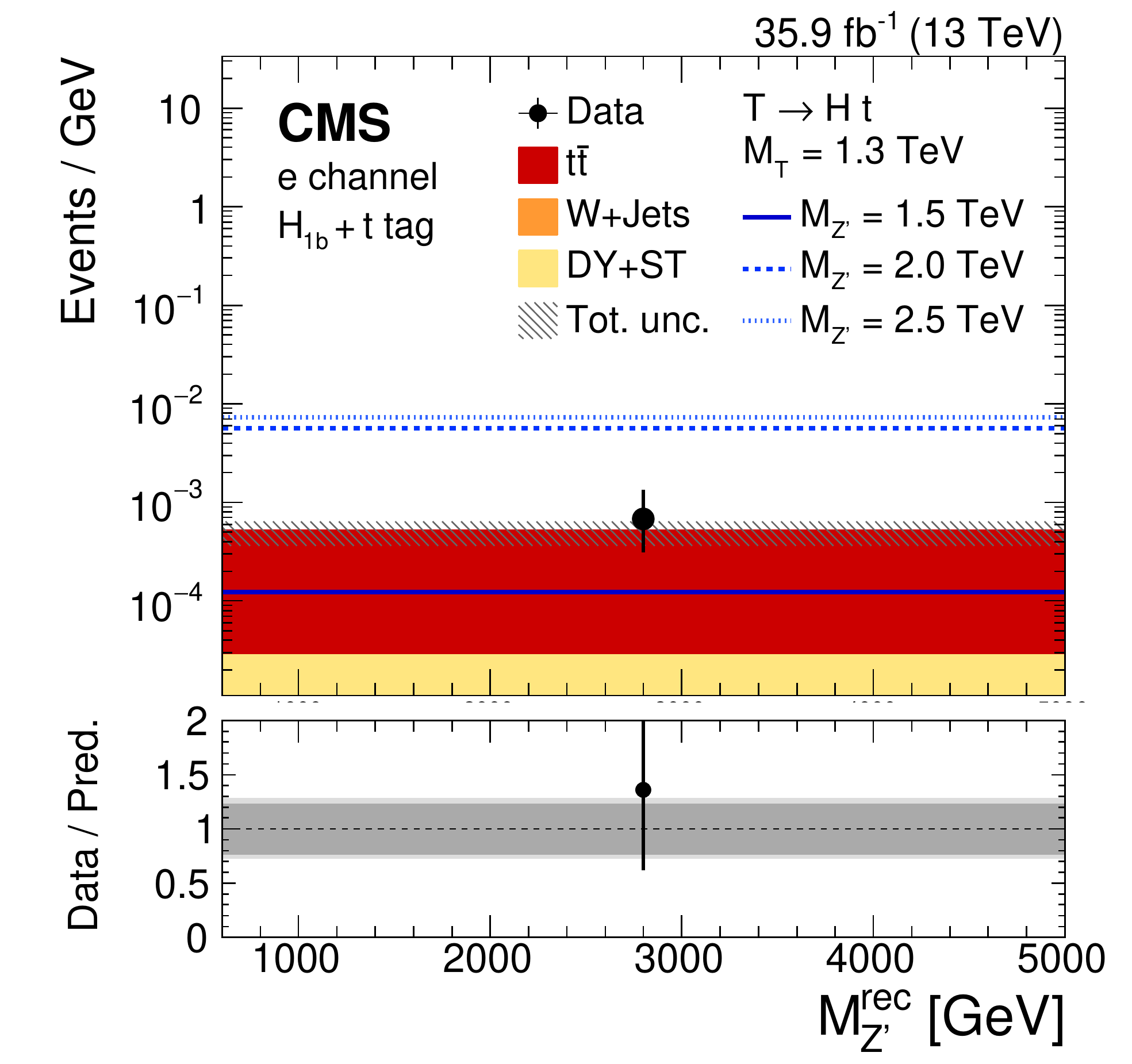}
\includegraphics[width=0.39\textwidth]{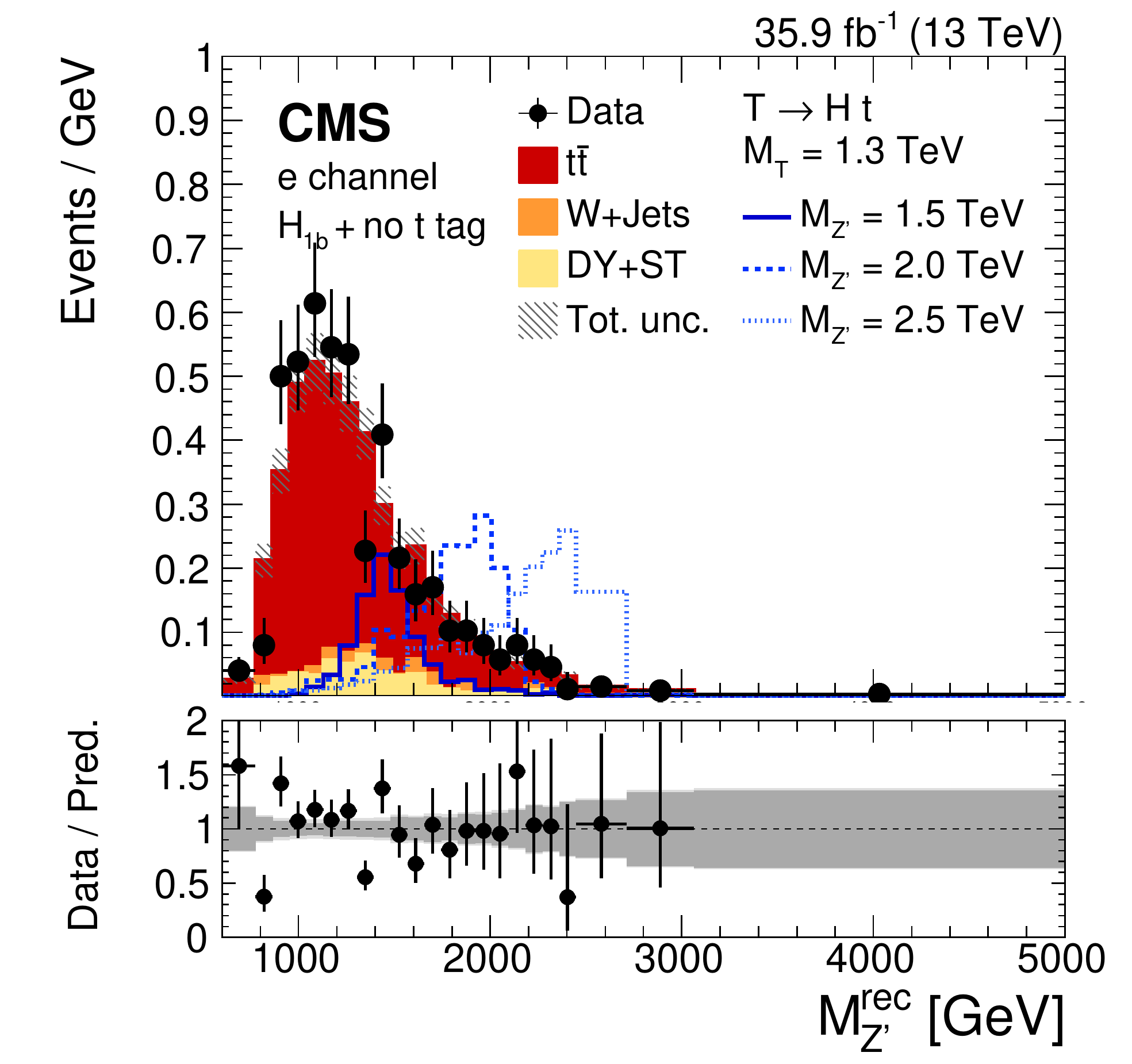} \\
\includegraphics[width=0.39\textwidth]{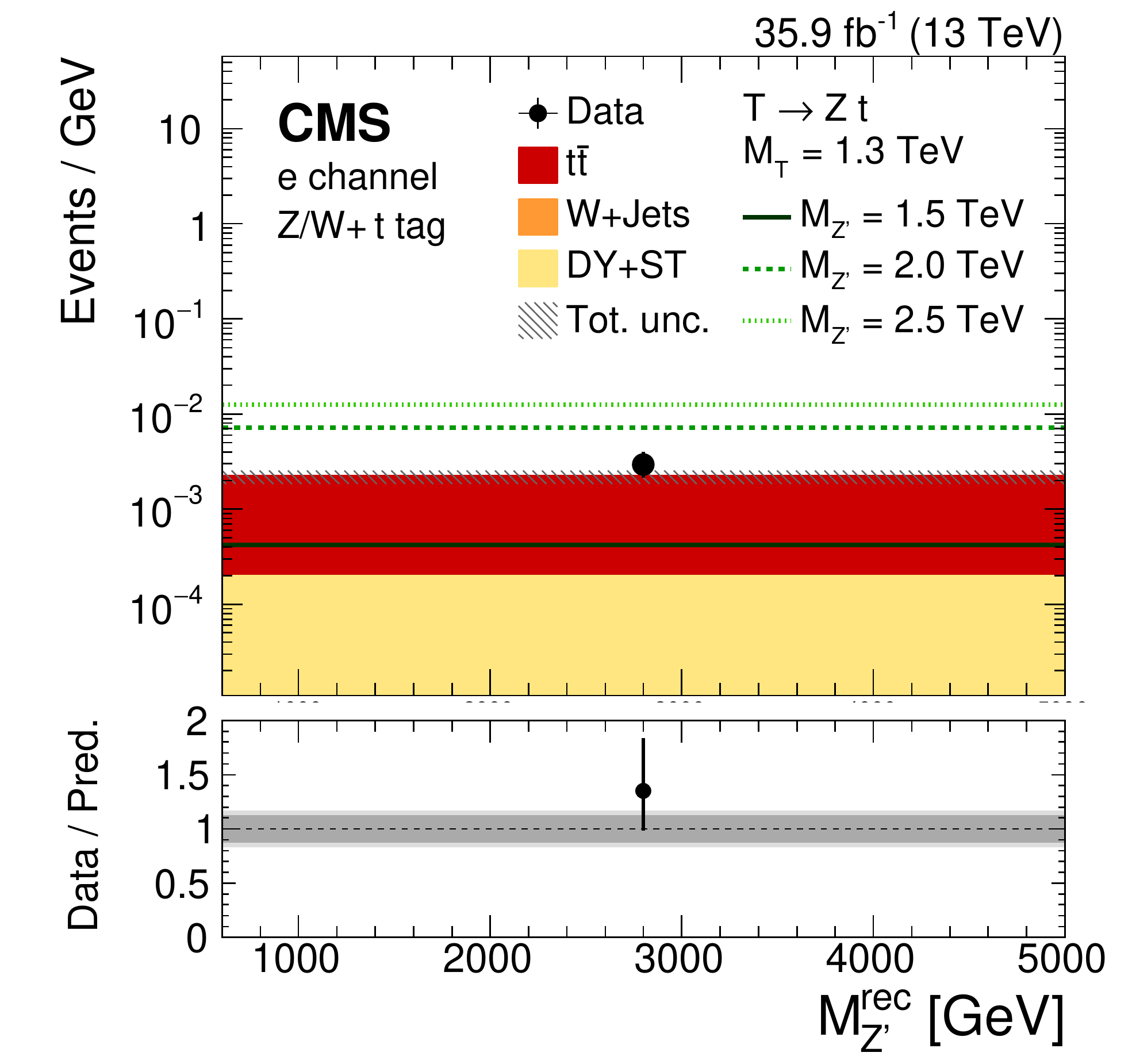}
\includegraphics[width=0.39\textwidth]{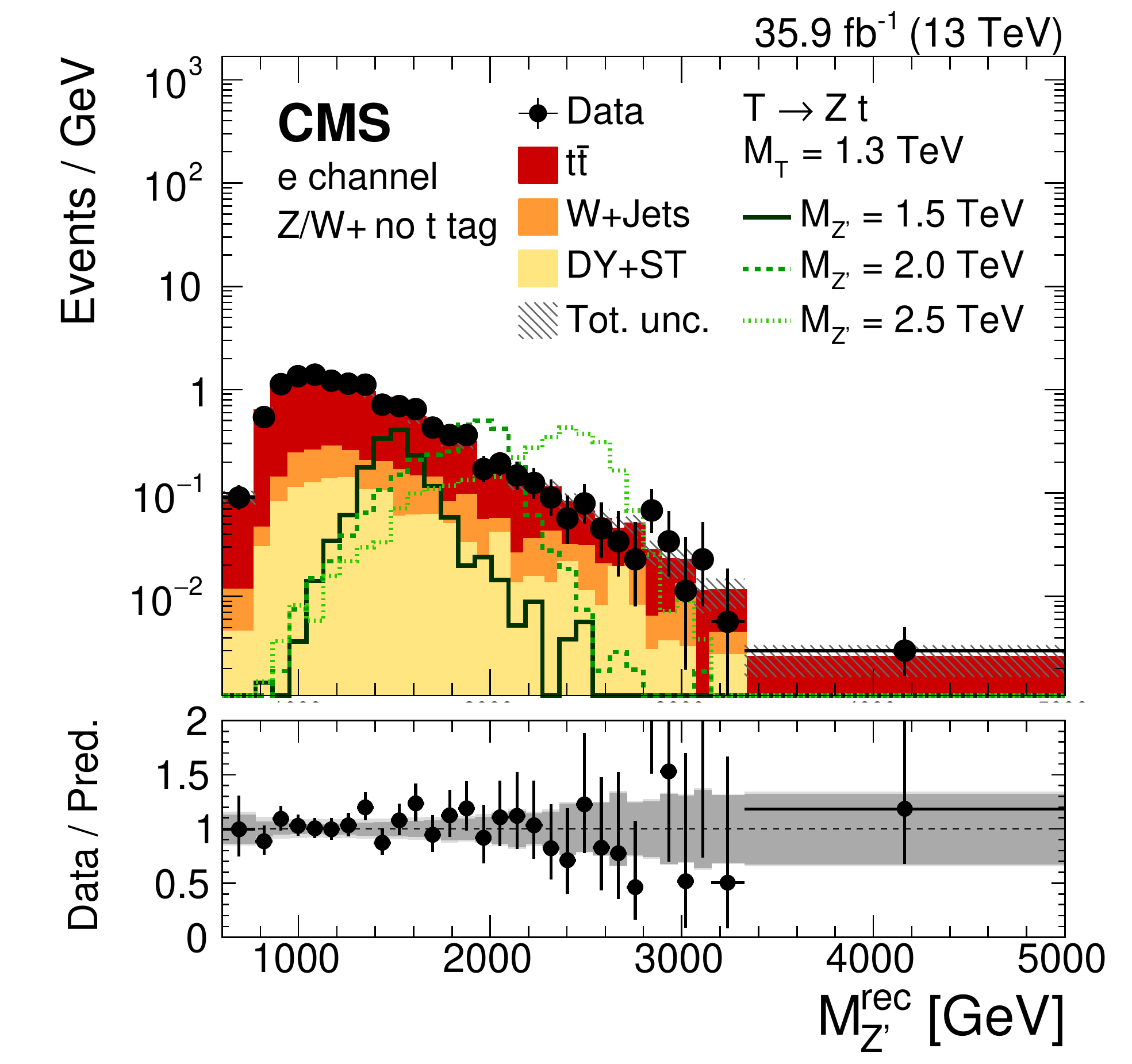}
\caption{Distribution of the reconstructed $\cPZpr$ resonance mass after the
  full selection in the {\ejets} channel for the data, the expected SM
  background, and for the signal with different $\cPZpr$ masses for a fixed
  $\Tp$ mass of 1.3\TeV. In the left (right) column the results in the
  top tag (no top tag) category are shown. Different rows display the
  distributions of events accepted by different taggers as well as the
  signal for the respective $\Tp$ decays: $\Hbb$ tagger and
  $\Tht$ decay (upper), $\Hb$ tagger and $\Tht$ decay (middle), and $\ZW$ tagger and $\Tzt$ decay (lower).
  The signal histograms correspond to a nominal cross section $\sigma(\cPZpr \to \cPqt\Tp)$ of 1\pb.
  The lower panel shows the ratio of data to predicted background. Here the darker grey band indicates the statistical uncertainty, whilst the lighter grey band shows the combined statistical and systematic uncertainty.}
\label{fig:final_elec}
\end{figure*}

The final reconstructed $\cPZpr$ invariant mass distribution is shown in each of the various categories in Fig.~\ref{fig:final_muon} for the muon channel, and in Fig.~\ref{fig:final_elec} for the electron channel.
The smaller number of events in the electron channel is due to the higher electron trigger threshold in comparison to the muon trigger threshold.
The binning in the figures is chosen such that the statistical uncertainty of the background
MC simulation in each bin does not exceed 30\%, leading to some categories represented by
a single bin only.
As a consequence of the higher electron trigger threshold, the categories ``$\ZW$ with top tag''
and ``$\Hbb$ with no top tag'' contain significantly fewer events
than the corresponding categories in the muon channel, and are thus only represented by a single
bin as well.

A binned likelihood combining the reconstructed $\cPZpr$ invariant mass distributions in all categories and channels is constructed to compare the signal and SM background hypotheses.
A Poisson probability is calculated in each bin of the mass distribution for each category in each channel.
The uncertainty due to the limited number of events in the templates is taken into account using a simplified Barlow--Beeston method that defines one additional nuisance parameter with a Gaussian distribution for each bin~\cite{barlow_beeston}.
The systematic uncertainties are taken into account as nuisance parameters in the likelihood.
For each systematic uncertainty that affects the shape of the reconstructed $\cPZpr$ mass, an interpolation between the nominal template and the shifted template with a Gaussian prior is performed.
Systematic uncertainties that affect only the normalization are taken into account as nuisance parameters with log-normal priors.
The likelihood is maximized with respect to these parameters.
The parameters representing the Poisson means of the signal strength and the background processes are determined in a maximum likelihood fit to the data, using a flat prior for the signal strength.

No significant excess of events in the data over the expectation from SM backgrounds was found for the
ranges of $\cPZpr$ and $\Tp$ masses considered.
A Bayesian calculation with priors known to yield good frequentist properties~\cite{bayes,PhysRevD.98.030001,Demortier:2010sn} is used to derive 95\% \CL upper limits on the product of the cross section and branching fraction, $\sigma(\pp \to \cPZpr \to \cPqt\Tp) \mathcal{B}(\Tp \to \PH\cPqt, \PZ\cPqt, \PW\cPqb)$, for a heavy resonance $\cPZpr$ decaying into a top quark and a vector-like quark $\Tp$.
The calculation is implemented in the \textsc{Theta} software package~\cite{theta}.
The median of the distribution of the upper limits at 95\% \CL in the pseudo-experiments and the central 68\% (95\%) interval define the expected upper limit and the 1 (2) standard deviation band, respectively.

\begin{figure*}[htbp!]
\centering
\includegraphics[width=0.47\textwidth]{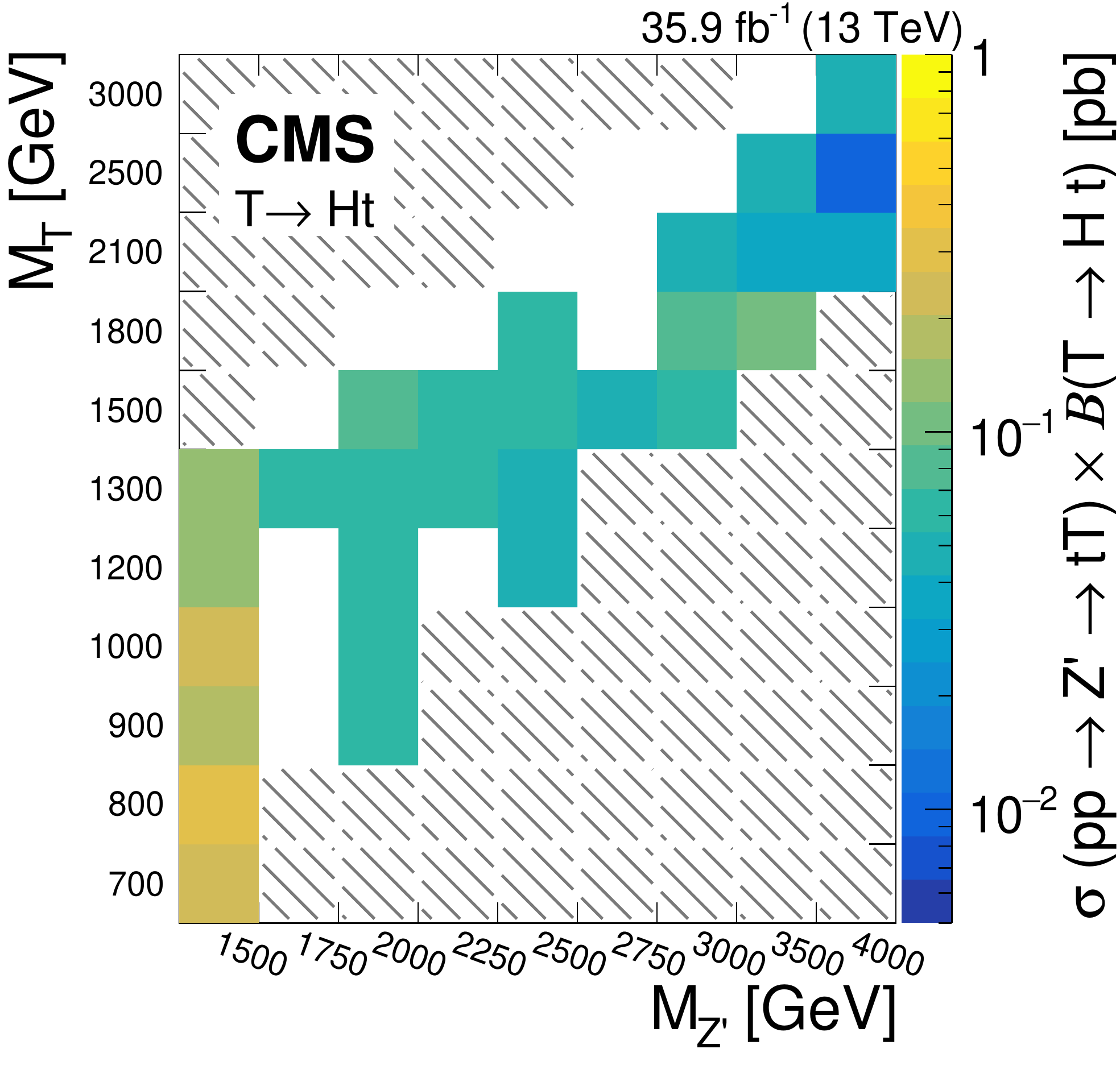}
\hfill
\includegraphics[width=0.47\textwidth]{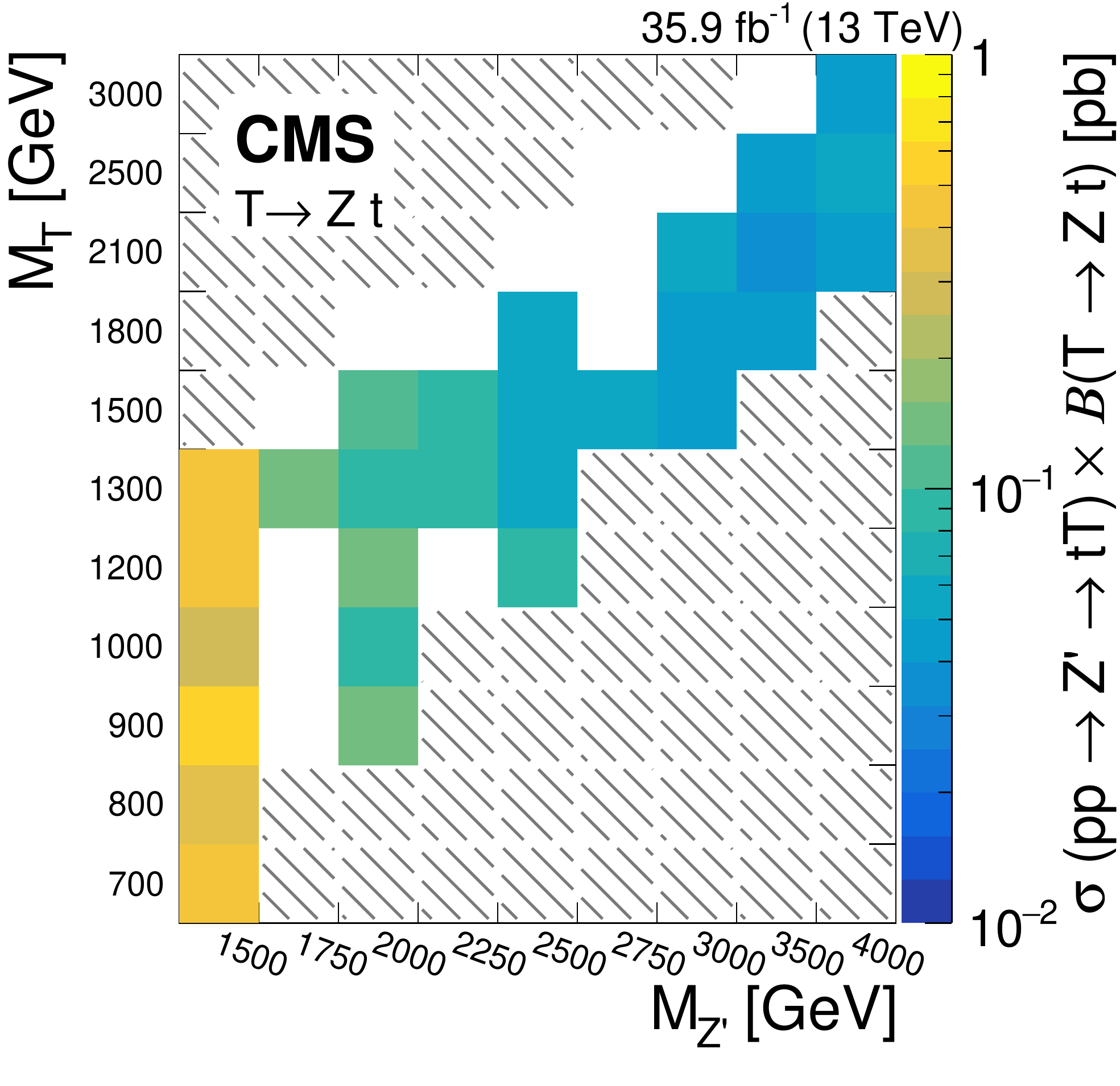} \\
\includegraphics[width=0.47\textwidth]{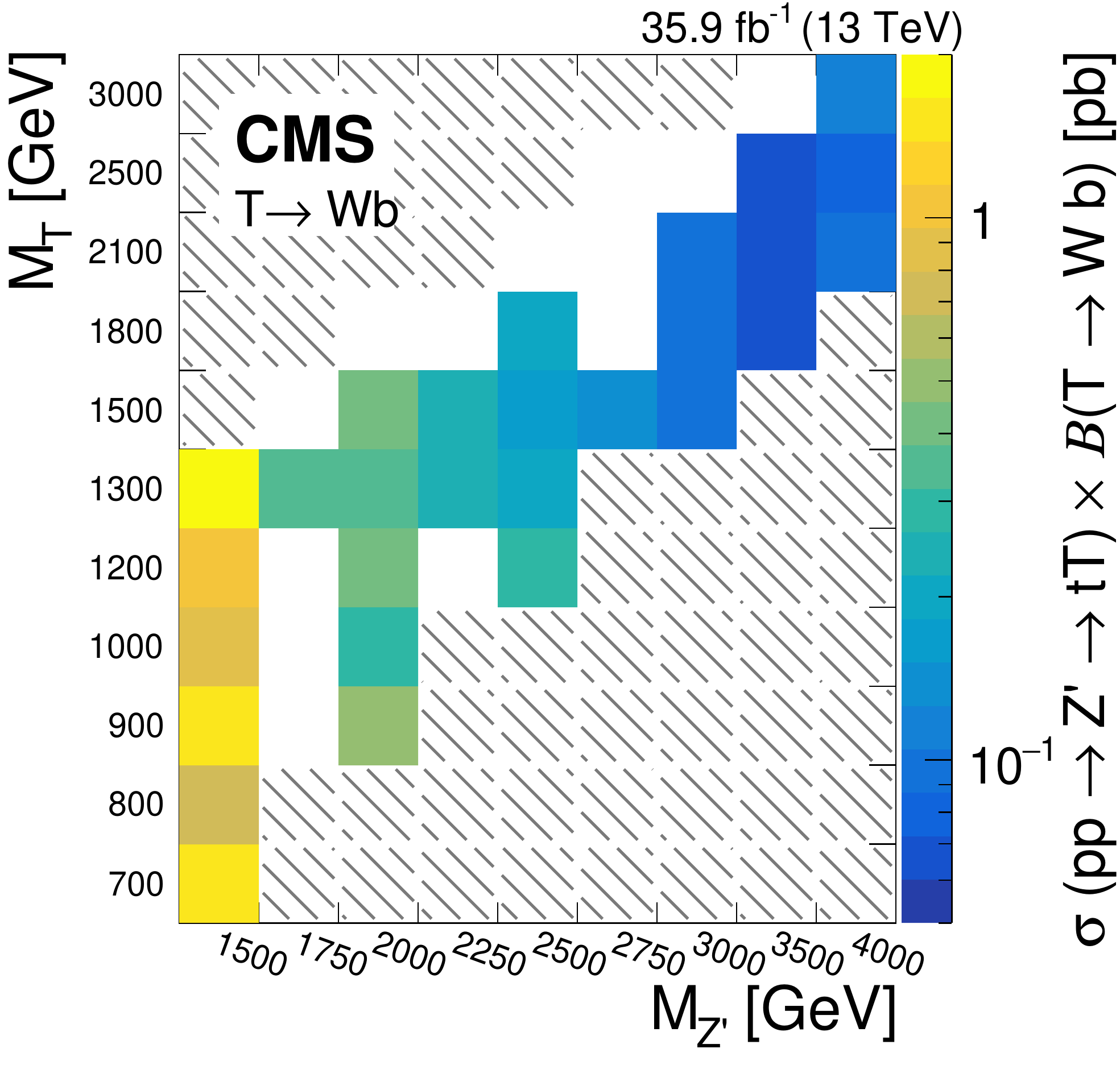}
\caption{Observed exclusion limits at 95\% \CL on the production cross section for various
  ($\MZp, \MTp$) combinations for the decay channels $\Tht$ (upper left), $\Tzt$ (upper right), and
  $\Twb$ (lower).
  The hatched area in the upper left indicates the region where the $\ZtT$ decay is kinematically forbidden,
  while in the lower right $\ZtT$ is suppressed by the preferred $\cPZpr \to \Tp\Tp$ mode.
  White areas indicate regions where signal samples have not been generated.}
\label{fig:limits_massmass}
\end{figure*}

Figure~\ref{fig:limits_massmass} shows observed limits as a function of
$\cPZpr$ mass, $\Tp$ mass, and $\Tp$ decay mode.
The limits are obtained using only decays to the indicated decay mode.
Limits are shown for combinations of $\cPZpr$ and $\Tp$ masses
where the decay $\ZtT$ is kinematically allowed and the
decay $\cPZpr \to \Tp\Tp$ is kinematically forbidden.

\begin{figure*}[htbp!]
\centering
\includegraphics[width=0.49\textwidth]{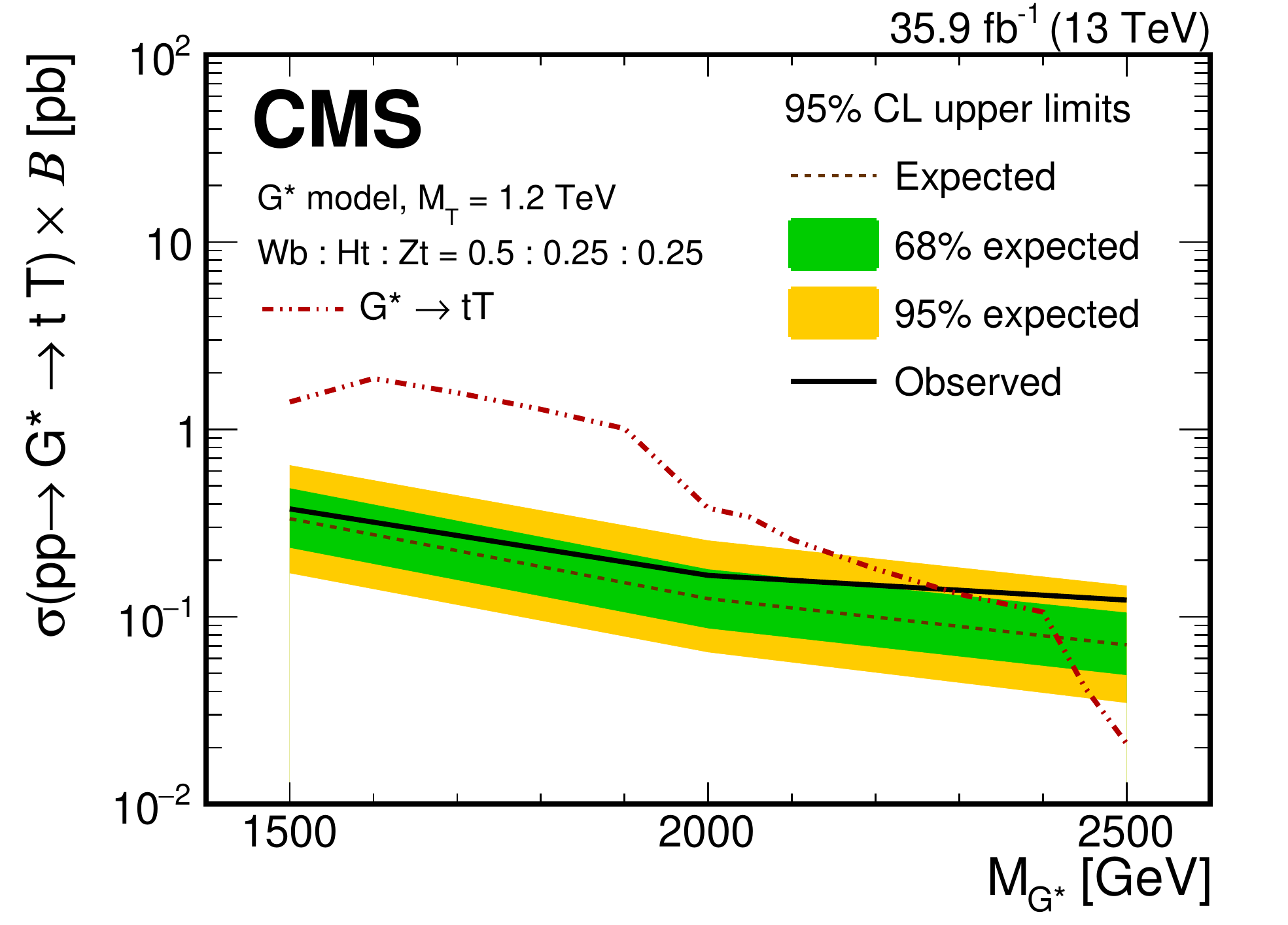}
\includegraphics[width=0.49\textwidth]{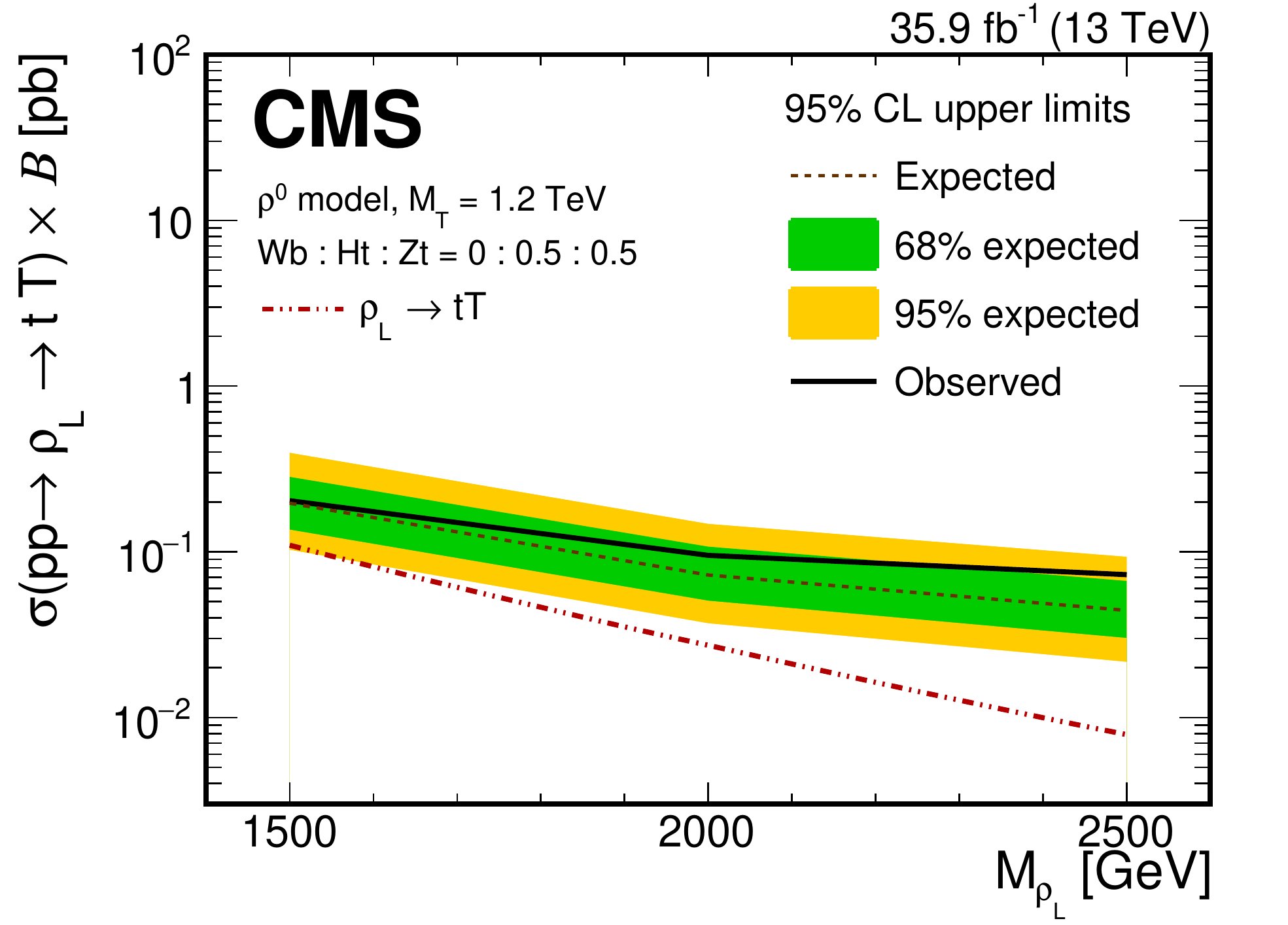} \\
\includegraphics[width=0.49\textwidth]{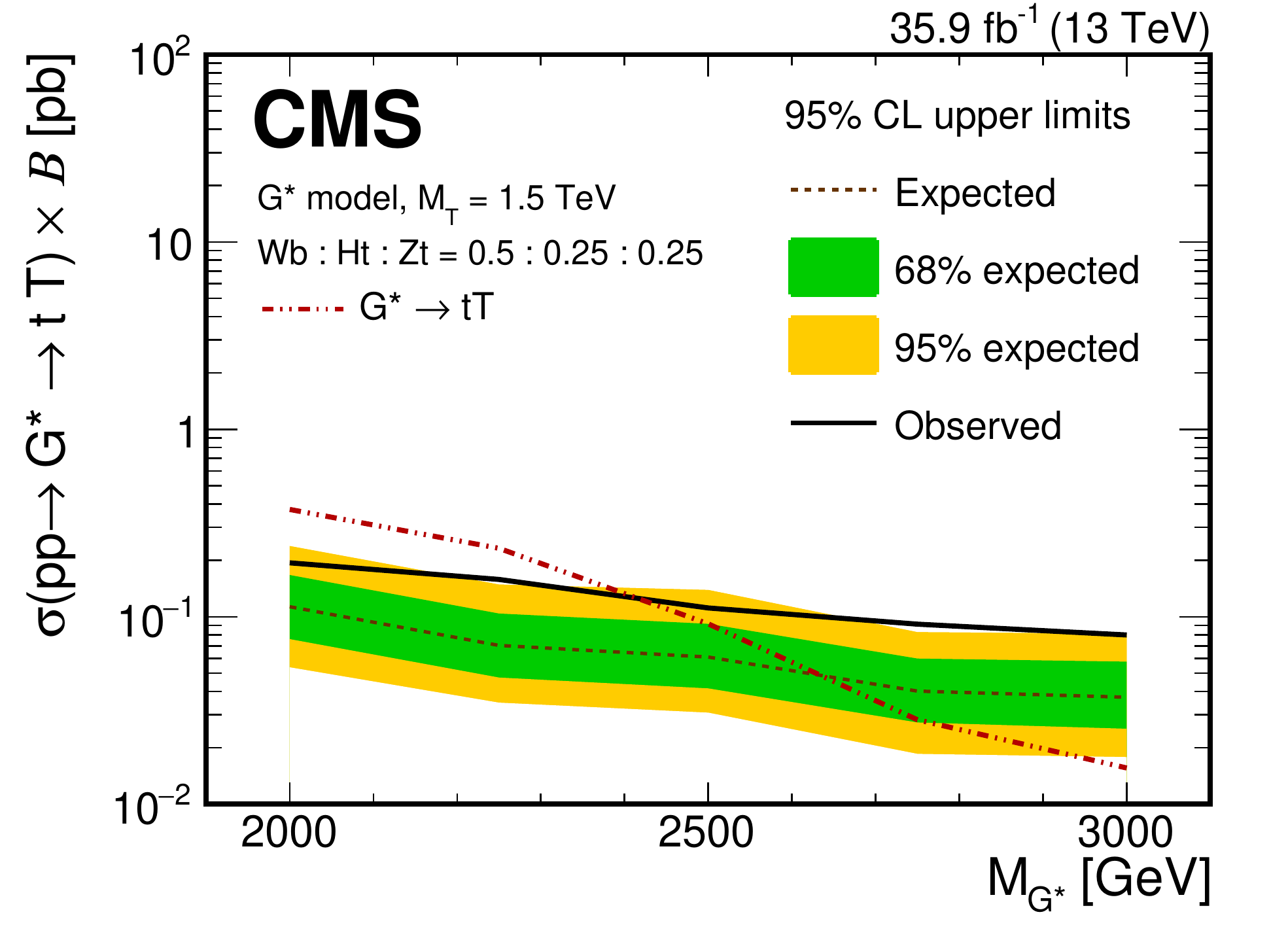}
\includegraphics[width=0.49\textwidth]{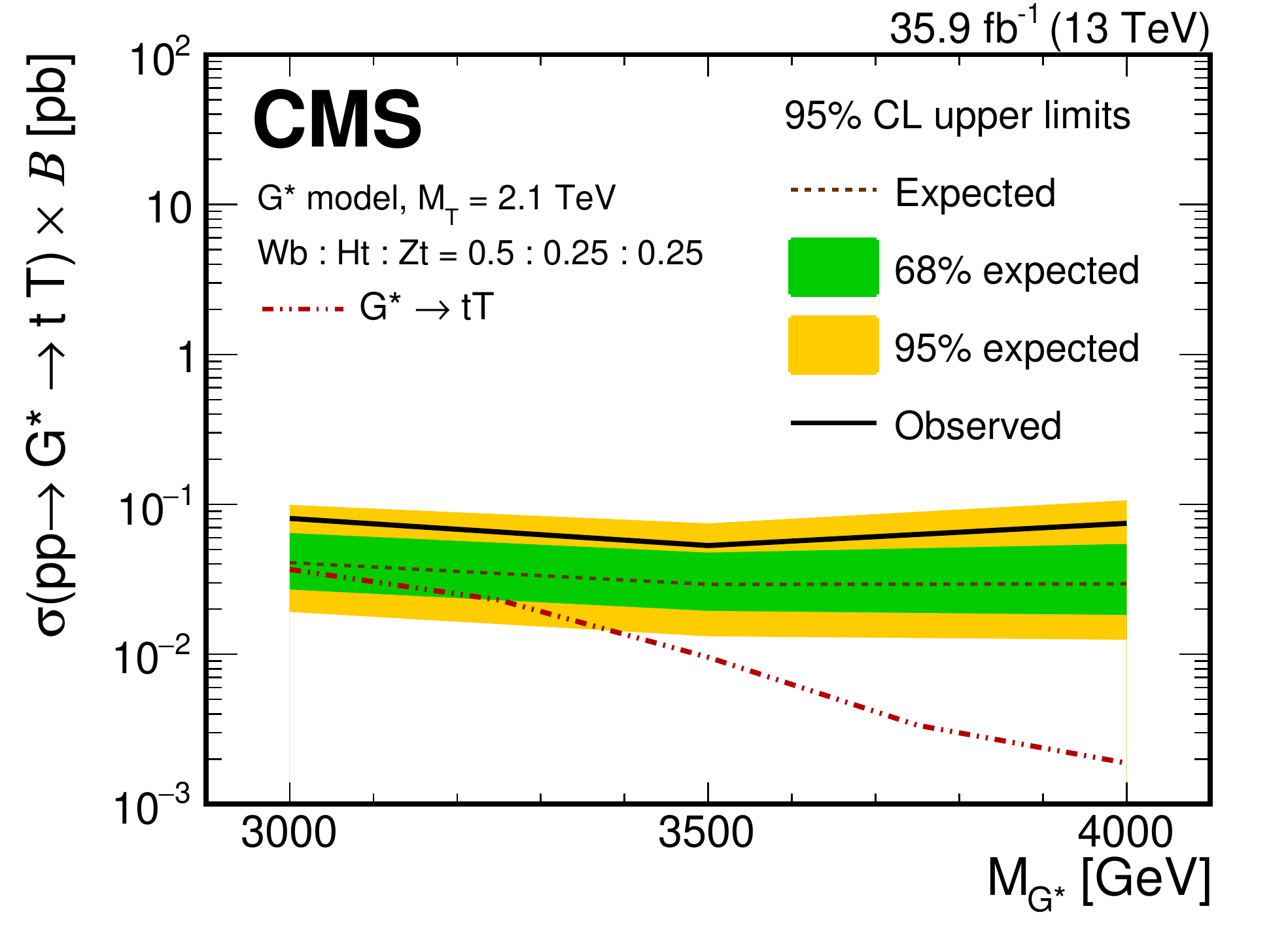}
\caption{Exclusion limits at 95\% \CL on the product of the cross
  section and branching fraction
  for three $\Tp$ masses of 1.2\TeV (upper row),
  1.5\TeV (lower left), and 2.1\TeV (lower right), as a function of the resonance mass.
  The branching fraction is defined as
  $\mathcal{B}= \mathcal{B}(\Twb) + \mathcal{B}(\Tht) + \mathcal{B}(\Tzt)$.
  Observed and expected limits are compared to the
  predictions from two different theory benchmark models:
  the $\gstar$ model (upper left and lower row),
  and the left-handed $\rhol$ in the $\rhoz$ model (upper right).
  }
\label{fig:limits_mass}
\end{figure*}

Figure~\ref{fig:limits_mass} shows limits on the product of the cross sections and branching fractions
$\mathcal{B} = \mathcal{B}(\Twb) + \mathcal{B}(\Tht) + \mathcal{B}(\Tzt)$, for fixed
ratios of the individual branching fractions.
The upper row in Fig.~\ref{fig:limits_mass} shows the limit
as a function of the $\cPZpr$ mass for a fixed $\Tp$ mass of 1.2\TeV.
The upper left plot compares the limit with a prediction from the $\gstar$ model,
showing that $\gstar$ masses between 1.5 and 2.3\TeV are excluded by this search,
for a $\Tp$ mass of 1.2\TeV.
The decrease in the predicted $\gstar$ cross section at a mass of approximately 2\TeV is due to the custodian VLQ $\Tp_{5/3}$ with a mass of 1\TeV, such that the $\gstar \to \Tp_{5/3} \Tp_{5/3}$ decay mode then becomes kinematically viable.
At a mass of 2.4\TeV there is another decrease in the predicted cross section due to the availability of the $\gstar \to \Tp\, \Tp$ decay. This has the effect of drastically increasing the width of the $\gstar$, and also lowering the branching ratio (and hence predicted cross section) for the decay mode $\gstar \to \cPqt\Tp$.
In the upper right plot of Fig.~\ref{fig:limits_mass}, the limit is compared with a prediction for the left-handed $\rhol$ from the $\rhoz$ model, showing that this search is not sensitive to this model.
The lower row of plots in Fig.~\ref{fig:limits_mass} shows the observed and expected limits in the context of the $\gstar$ model for two other $\Tp$ masses.
For a $\Tp$ mass of 1.5\TeV (lower left), $\gstar$ masses between 2.0 and 2.4\TeV are excluded by this search, whilst for a $\Tp$ mass of 2.1\TeV (lower right) this analysis is not able to exclude the model scenario.

\begin{figure}[htb!]
\centering
\includegraphics[width=0.49\textwidth]{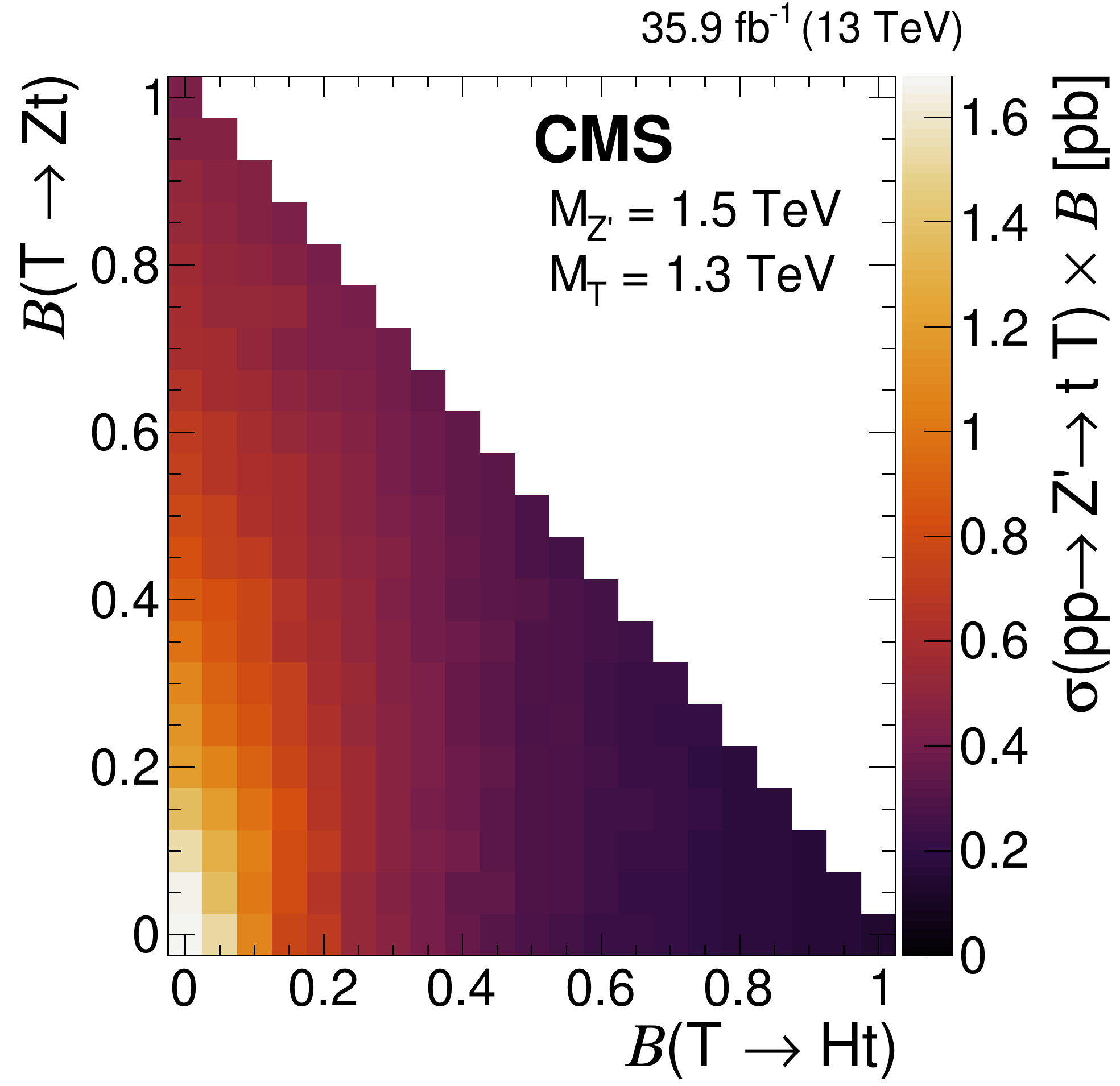}
\caption{Model-independent observed exclusion limits at 95\% \CL on the product of the cross
section and branching fraction
$\mathcal{B}= \mathcal{B}(\Twb) + \mathcal{B}(\Tht) + \mathcal{B}(\Tzt)$
for an example mass configuration of $\MZp=1.5$\TeV and $\MTp=1.3$\TeV as a function of the
branching fractions $\mathcal{B}(\Tht)$ and $\mathcal{B}(\Tzt)$.
\label{fig:limits_brbr}}
\end{figure}

Finally, Fig.~\ref{fig:limits_brbr} shows the observed limits on the product of the cross section and branching fraction as a function of the branching fractions $\mathcal{B}(\Tht)$ and $\mathcal{B}(\Tzt)$ for a $\cPZpr$ mass of 1.5\TeV and a $\Tp$ mass of 1.3\TeV, demonstrating the dependence of the limit on both branching fractions.

The expected and observed limits are 30\% weaker for a $\cPZpr$ width of 30\% when compared to a
width of 1\% for the mass range $1.5 < \MZp < 2.5\TeV$.
However for $\MZp > 3\TeV$ there is no
significant difference between the limits obtained with each of the widths.

\section{Summary}
\label{sec:summary}
A search for a heavy spin-1 resonance $\cPZpr$ decaying to a standard model top quark and a
vector-like quark partner $\Tp$ has been presented.
The data used in this search were recorded with the CMS detector at the LHC at $\sqrt{s} = 13\TeV$
and correspond to an integrated luminosity of $\totlumi$.
The analysis is primarily optimised to study the decay modes of the vector-like quark to a Higgs
boson and a top quark ($\Tht$), and to a \PZ boson and a top quark ($\Tzt$),
although the decay to a $\PW$ boson and a bottom quark ($\Twb$) is also considered.
This is the first direct search for the decay $\ZtT \to \cPqt\PH\cPqt$.
No significant excess of events over the expectation from standard model backgrounds is found.
Limits on the production cross section are presented for a narrow $\cPZpr$ resonance in the mass range from 1.5 to 4.0\TeV and a narrow $\Tp$ resonance in the mass range from 0.7 to 3.0\TeV.
Interpretation of these limits within the context of the $\gstar$ benchmark model results in
the exclusion of $\gstar$ resonance masses in the range from 1.5 to 2.3\TeV and from 2.0 to 2.4\TeV,
for a $\Tp$ mass of 1.2 and 1.5\TeV, respectively.
The presented limits are the most stringent to date for the decay mode $\ZtT \to \cPqt\PH\cPqt$.

\begin{acknowledgments}
We congratulate our colleagues in the CERN accelerator departments for the excellent performance of the LHC and thank the technical and administrative staffs at CERN and at other CMS institutes for their contributions to the success of the CMS effort. In addition, we gratefully acknowledge the computing centres and personnel of the Worldwide LHC Computing Grid for delivering so effectively the computing infrastructure essential to our analyses. Finally, we acknowledge the enduring support for the construction and operation of the LHC and the CMS detector provided by the following funding agencies: BMBWF and FWF (Austria); FNRS and FWO (Belgium); CNPq, CAPES, FAPERJ, FAPERGS, and FAPESP (Brazil); MES (Bulgaria); CERN; CAS, MoST, and NSFC (China); COLCIENCIAS (Colombia); MSES and CSF (Croatia); RPF (Cyprus); SENESCYT (Ecuador); MoER, ERC IUT, and ERDF (Estonia); Academy of Finland, MEC, and HIP (Finland); CEA and CNRS/IN2P3 (France); BMBF, DFG, and HGF (Germany); GSRT (Greece); NKFIA (Hungary); DAE and DST (India); IPM (Iran); SFI (Ireland); INFN (Italy); MSIP and NRF (Republic of Korea); MES (Latvia); LAS (Lithuania); MOE and UM (Malaysia); BUAP, CINVESTAV, CONACYT, LNS, SEP, and UASLP-FAI (Mexico); MOS (Montenegro); MBIE (New Zealand); PAEC (Pakistan); MSHE and NSC (Poland); FCT (Portugal); JINR (Dubna); MON, RosAtom, RAS, RFBR, and NRC KI (Russia); MESTD (Serbia); SEIDI, CPAN, PCTI, and FEDER (Spain); MOSTR (Sri Lanka); Swiss Funding Agencies (Switzerland); MST (Taipei); ThEPCenter, IPST, STAR, and NSTDA (Thailand); TUBITAK and TAEK (Turkey); NASU and SFFR (Ukraine); STFC (United Kingdom); DOE and NSF (USA).

\hyphenation{Rachada-pisek} Individuals have received support from the Marie-Curie programme and the European Research Council and Horizon 2020 Grant, contract No. 675440 (European Union); the Leventis Foundation; the A.P.\ Sloan Foundation; the Alexander von Humboldt Foundation; the Belgian Federal Science Policy Office; the Fonds pour la Formation \`a la Recherche dans l'Industrie et dans l'Agriculture (FRIA-Belgium); the Agentschap voor Innovatie door Wetenschap en Technologie (IWT-Belgium); the F.R.S.-FNRS and FWO (Belgium) under the ``Excellence of Science -- EOS" -- be.h project n.\ 30820817; the Ministry of Education, Youth and Sports (MEYS) of the Czech Republic; the Lend\"ulet (``Momentum") Programme and the J\'anos Bolyai Research Scholarship of the Hungarian Academy of Sciences, the New National Excellence Program \'UNKP, the NKFIA research grants 123842, 123959, 124845, 124850, and 125105 (Hungary); the Council of Science and Industrial Research, India; the HOMING PLUS programme of the Foundation for Polish Science, cofinanced from European Union, Regional Development Fund, the Mobility Plus programme of the Ministry of Science and Higher Education, the National Science Center (Poland), contracts Harmonia 2014/14/M/ST2/00428, Opus 2014/13/B/ST2/02543, 2014/15/B/ST2/03998, and 2015/19/B/ST2/02861, Sonata-bis 2012/07/E/ST2/01406; the National Priorities Research Program by Qatar National Research Fund; the Programa Estatal de Fomento de la Investigaci{\'o}n Cient{\'i}fica y T{\'e}cnica de Excelencia Mar\'{\i}a de Maeztu, grant MDM-2015-0509 and the Programa Severo Ochoa del Principado de Asturias; the Thalis and Aristeia programmes cofinanced by EU-ESF and the Greek NSRF; the Rachadapisek Sompot Fund for Postdoctoral Fellowship, Chulalongkorn University and the Chulalongkorn Academic into Its 2nd Century Project Advancement Project (Thailand); the Welch Foundation, contract C-1845; and the Weston Havens Foundation (USA). \end{acknowledgments}
\clearpage
\bibliography{auto_generated}

\cleardoublepage \appendix\section{The CMS Collaboration \label{app:collab}}\begin{sloppypar}\hyphenpenalty=5000\widowpenalty=500\clubpenalty=5000\vskip\cmsinstskip
\textbf{Yerevan Physics Institute, Yerevan, Armenia}\\*[0pt]
A.M.~Sirunyan, A.~Tumasyan
\vskip\cmsinstskip
\textbf{Institut f\"{u}r Hochenergiephysik, Wien, Austria}\\*[0pt]
W.~Adam, F.~Ambrogi, E.~Asilar, T.~Bergauer, J.~Brandstetter, M.~Dragicevic, J.~Er\"{o}, A.~Escalante~Del~Valle, M.~Flechl, R.~Fr\"{u}hwirth\cmsAuthorMark{1}, V.M.~Ghete, J.~Hrubec, M.~Jeitler\cmsAuthorMark{1}, N.~Krammer, I.~Kr\"{a}tschmer, D.~Liko, T.~Madlener, I.~Mikulec, N.~Rad, H.~Rohringer, J.~Schieck\cmsAuthorMark{1}, R.~Sch\"{o}fbeck, M.~Spanring, D.~Spitzbart, A.~Taurok, W.~Waltenberger, J.~Wittmann, C.-E.~Wulz\cmsAuthorMark{1}, M.~Zarucki
% ...
\vskip\cmsinstskip
% ...
1:  Also at Vienna University of Technology, Vienna, Austria\\
% ...
\end{sloppypar}
\end{document}